\documentclass[twocolumn]{aastex631}

\usepackage{comment}
\usepackage{amsmath}

\shorttitle{A Pathway for Collisional Planetesimal Growth}
\shortauthors{Yunerman et al.}

\graphicspath{{./}{figs/}}

\begin{document}

\title{A Pathway for Collisional Planetesimal Growth in the Ice-Dominant Regions of Protoplanetary Disks}

\author[0009-0008-4937-3314]{Elizabeth Yunerman}
\affiliation{Center for Astrophysics $|$ Harvard \& Smithsonian, 60 Garden St, Cambridge, MA 02138, USA}

\author[0000-0002-4250-0957]{Diana Powell}
\affiliation{Department of Astronomy \& Astrophysics, University of Chicago, 5640 S Ellis Ave, Chicago, IL 60637, USA}

\author[0000-0001-5061-0462]{Ruth Murray-Clay}
\affiliation{Department of Astronomy and Astrophysics, University of California, Santa Cruz, CA 95064, USA}

%%%%%%%%%%%%%%%%%%%%%%%%%%%%%%%%%%%%%%%%%%%%%
%%%%%%%%%%%%%%%%%% ABSTRACT %%%%%%%%%%%%%%%%%
%%%%%%%%%%%%%%%%%%%%%%%%%%%%%%%%%%%%%%%%%%%%%

\begin{abstract}
We present a semi-analytic model for the growth, drift, desorption, and fragmentation of millimeter- to meter-sized particles in protoplanetary disks. Fragmentation occurs where particle collision velocities exceed critical fragmentation velocities.  Using this criterion, we produce fragmentation regions in disk orbital radius--particle size phase space for particles with a range of material properties, structures, and compositions (including SiO$_2$, Mg$_2$SiO$_4$, H$_2$O, CO$_2$, and CO).  For reasonable disk conditions, compact aggregate H$_2$O, CO$_2$, and CO ice particles do not reach destructive relative velocities and are thus not likely to undergo collisional fragmentation. Uncoated silicate particles are more susceptible to collisional destruction and are expected to fragment in the inner disk, consistent with previous work. We then calculate the growth, drift, and sublimation of small particles, initially located in the outer disk. We find that ice-coated particles can avoid fragmentation as they grow and drift inward under a substantial range of disk conditions as long as the particles are aggregates composed of 0.1 $\mu$m-sized monomers. Such particles may undergo runaway growth in disk regions abundant in H$_2$O or CO$_2$ ice depending on the assumed disk temperature structure. These results indicate that icy collisional growth to planetesimally-relevant sizes may happen efficiently throughout a disk's lifetime, and is particularly robust at early times when the disk's dust-to-gas ratio is comparable to that of the interstellar medium.

\end{abstract}


%%%%%%%%%%%%%%%%%%%%%%%%%%%%%%%%%%%%%%%%%%%%%
%%%%%%%%%%%%%%% INTRODUCTION %%%%%%%%%%%%%%%%
%%%%%%%%%%%%%%%%%%%%%%%%%%%%%%%%%%%%%%%%%%%%%

\section{Introduction} \label{sec:intro}
In the classical picture of particle evolution during the initial stages of planet formation, collisional growth of solids to the planetesimal size scale is thought to be prevented by several limitations that are jointly known as the “meter-size barrier". While other processes may be relevant to this problem (discussed in Section \ref{sec:disc_other_barriers}), the classical meter-size barrier is the barrier to continued particle growth beyond millimeter to sub-kilometer scales due to either collisional fragmentation or particle drift (see \cite{Chiang2010,Birnstiel2016} for reviews). Fragmentation occurs when particles collide at relative velocities that are energetic enough to exceed the surface energy keeping the particles intact. The  particle--particle relative velocities are maximized in the meter-size regime at separations of $\sim$1AU from the host star, resulting in potentially destructive collisions \citep{Blum2000a}. These particles also become large enough to begin to decouple from the dynamics of the background gas, such that significant gas drag causes particles to rapidly drift inwards towards the system's host star. The effects of both collisional fragmentation and inward drift prevent particles in this regime from collisionally growing to large sizes. 

Given the challenges of continued collisional growth past the meter-size barrier, alternative models for the formation of $\sim$1-100km planetesimals have been proposed. Resonant Drag Instabilities (RDIs), such as the streaming instability, have been shown to create particle overdensities which can directly gravitationally collapse into planetesimals \citep[e.g.,][]{Goldreich1973,Youdin2002,Youdin2005,Johansen2006,Chiang2010,Simon2016,Squire2018,Gerbig2020}. While models of unstable RDIs successfully produce planetesimals, they rely on strict initial conditions that may not represent commonly occurring disk conditions. 

Enough uncertainty remains, however, in the behavior of particle collisions in disks that collisional growth may still be a viable planetesimal formation mechanism. While some of this uncertainty comes from the protoplanetary disk conditions during formation, crucially, there are significant uncertainties in the material properties of solids. These include, but are not limited to, the ice-coated particles composition, structure (i.e., non-aggregate vs. aggregate with small or large monomers, sintered vs. unsintered monomer connections, crystalline vs. amorphous, etc.), physical surface processes between silica-ice and ice-ice boundaries \citep{Fogarty2010,Nietiadi2020}, and compaction behavior of porous particles upon impact \citep{Paszun2009,Krijt2015}.

Particles beyond the water ice line are expected to be coated in ices with material properties that differ from silicate grains, thus altering the particles' strength in withstanding fragmentation. For example, several laboratory studies have shown that the critical velocity for H$_2$O ice, and grains which are coated in ice may be higher than that for silicates, indicating that ice-coated particles may be more robust against collisional fragmentation (\citealt{Poppe2000,Blum2008,Gundlach2011,Gundlach2015,Musiolik2016a,Musiolik2016b}; see Section \ref{sec:materialprops} for further discussion). This is thought to be largely due to the increased surface energy of H$_2$O ice, particularly at temperatures near the ice line \citep{Gundlach2018,Musiolik2019}. \cite{Musiolik2016a} investigate CO$_2$ ice particles and find that mixtures between CO$_2$ and H$_2$O can increase the overall sticking during collisions. In terms of particle structure, \cite{Kataoka2013} find that micron-sized icy dust aggregates may stick together forming fluffy planetesimals that are able to surpass the drift barrier without fully fragmenting, and numerical models by \cite{Wada2007,Wada2009b} reveal that icy dust aggregates are able to retain material during collisions of varying impact parameters and velocities. Collision-induced heating of the ice particles may also be an important factor that changes the surface layer physics from dry to wet \citep{Nietiadi2020}. Ultimately these studies demonstrate that particle composition (both in grain and ice mantles) and the corresponding particle material properties are of first order importance when modeling particle evolution in disks. 

In this work, we investigate the implications of a species-dependent fragmentation velocity on the growth and evolution of particles in disks. We build upon the critical fragmentation velocity frameworks from \cite{Wada2007,Wada2009b} and \cite{Stewart2009} for a variety of particle compositions, porosities, and material properties. This model calculates particle relative velocities based on the relevant drag regimes present throughout the disk (i.e., Epstein, Stokes, and ram pressure drag regimes), and finds regions of fragmentation as a function of particle size and orbital radius by comparing the relative velocities with the respective critical fragmentation velocities. 

We produce the expected fragmentation regions for SiO$_2$, Mg$_2$SiO$_4$, H$_2$O, CO$_2$, and CO solid particles in TW Hya and in the Minimum Mass Solar Nebula (MMSN). We investigate the consequences of assumptions regarding particle material properties and gas drag regimes on the likelihood of particle fragmentation and growth, and discuss whether particles are strongly affected by the classical meter-size barrier. We find that compact aggregate H$_2$O, CO$_2$, and CO ice particles in the outer regions of protoplanetary disks may not undergo collisional fragmentation. Particle drift and desorption of ice are incorporated in the model such that the radial particle evolution and growth is compared to the fragmentation regions. We describe favorable locations for continued particle growth in the disk typically $\sim$1-10 AU, either near the CO$_2$ ice line or exterior to the H$_2$O ice line depending on the assumed disk temperature profile. These regions may allow for efficient runaway solid growth that is not limited by fragmentation, drift, or desorption, where the classical meter-size barrier is unlikely to operate. 

In Section \ref{sec:materialprops} we discuss the relevant material properties that govern particle collisions and in Section \ref{sec:critical_velocity} derive the corresponding critical velocities used in this study. In Section \ref{sec:disk_params} we outline the surface density profile and temperature dependencies in protoplanetary disks used in the model. Particle relative velocities are then derived in Section \ref{sec:relative_velocity}. The resulting fragmentation region phase space is described in Section \ref{sec:frag regions}, with favorable regions for planetesimal formation detailed in Section \ref{favorable_regions}. Implications of these results are discussed in Section \ref{discussion}, including other potential barriers to particle growth. A summary and conclusion of this work can be found in Section \ref{summary}.

%%%%%%%%%%%%%%%%%%%%%%%%%%%%%%%%%%%%%%%%%%%%%
%%%%%%%%%%% PARTICLE PROPERTIES %%%%%%%%%%%%%
%%%%%%%%%%%%%%%%%%%%%%%%%%%%%%%%%%%%%%%%%%%%%

\section{Particle Properties and Growth} \label{sec:materialprops}

We model the fragmentation of key volatiles in protoplanetary disks including H$_2$O, CO$_2$, and CO, as well as two rock species that are only volatile in the very innermost high-temperature regions of the disk. While previous studies have frequently focused on the fragmentation of silicates, such as SiO$_2$, we note that in comets the majority of non-volatile solid material has a composition similar to that of Mg$_2$SiO$_4$ \citep[][]{Wooden2007}. We thus consider the fragmentation properties of both of these species. 

\begin{deluxetable*}{lllll} [!htb]
\tablecolumns{5}
\tablecaption{Species Material Properties \label{material_props}}
\tablehead{ % column headings
 \colhead{Species} &
 \colhead{Surface Energy [erg cm$^{-2}$]} &
 \colhead{Young's Modulus [GPa]} &
 \colhead{Poisson's Ratio} &
 \colhead{Material Density [g cm$^{-2}$]}
}
\startdata
SiO$_2$ & 150\tablenotemark{a} & 54\tablenotemark{b} & 0.17\tablenotemark{b} & 2.65\\
Mg$_2$SiO$_4$ & 436\tablenotemark{c} & 187.05\tablenotemark{d} & 0.24259\tablenotemark{d} & 3.27\\
H$_2$O\tablenotemark{b} & 100 & 7 & 0.25 & 0.92 \\ 
CO$_2$ & 60\tablenotemark{e} & 10.7\tablenotemark{f} & 0.26\tablenotemark{f} & 1.56\\
CO & 23.1\tablenotemark{g} &  7 & 0.25 & 0.87\tablenotemark{h}\\
\enddata 
\tablerefs{ \tablenotemark{a}\citet{Kimura2020} \tablenotemark{b}\citet{Wada2009b} \tablenotemark{c}\citet{Kozasa1989} \tablenotemark{d}\citet{Gaillac2016} \tablenotemark{e}\citet{Fritscher2022} \tablenotemark{f}\cite{Arakawa2021} \tablenotemark{g}\citet{Sprow1966}
\tablenotemark{h}\citet{Luna2022}}
\end{deluxetable*}
The following properties determine the strength of an ice-aggregate, thus determining the likelihood that a particle will fragment upon collision: surface energy ($\gamma$), Young's modulus ($E$), Poisson's ratio ($\nu$), and material density ($\rho_i$); material properties for each ice species are given in Table~\ref{material_props}. Young's modulus quantifies how deformable and stretchy a material is---the higher the value the less deformable the material \citep{Heindl1936}. Poisson's ratio quantifies how much a material will stretch in the perpendicular direction of the impacting load, typically ranging from 0 to 0.5 \citep{Greaves2011}. The size of the monomers composing the ice-aggregate also determines the overall strength and is discussed in context of the material properties below. We further include the binding energy of H$_2$O, CO$_2$, and CO in Table \ref{table: ice line_params} which, in this context, is the energy required to remove ice from the surface of a grain. In this work the binding energy determines the ice desorption rate and thus sets the ice line location for each species (see Section \ref{sec:disk_params}). A species' surface energy is the amount of energy per unit area required to make (or destroy) a surface, and is a crucial component in calculating the critical fragmentation velocity in our model.

\begin{deluxetable}{llll} [!htb]
\tablecolumns{4}
\tablecaption{Ice Line Molecular Properties \label{table: ice line_params}}
\tablehead{ % column headings
 \colhead{Molecular Species} &
 \colhead{$E_i/k$ [K]} &
 \colhead{$n_i\times 10^{-4}$ [cm$^{-3}$]}
}
\startdata
H$_2$O & 5800\tablenotemark{a} & 0.9\tablenotemark{c} \\
CO$_2$ & 2000\tablenotemark{b} & 0.3\tablenotemark{c} \\
CO & 850\tablenotemark{b} & 1.5\tablenotemark{c} \\
\enddata 
\tablerefs{ \tablenotemark{a}\citet{Sandford1988} \tablenotemark{b}\citet{Aikawa1996}
\tablenotemark{c}\citet{Pontoppidan2006}}
\end{deluxetable}

\begin{figure*}[!htb]
 \centering
 \includegraphics[width=12cm]{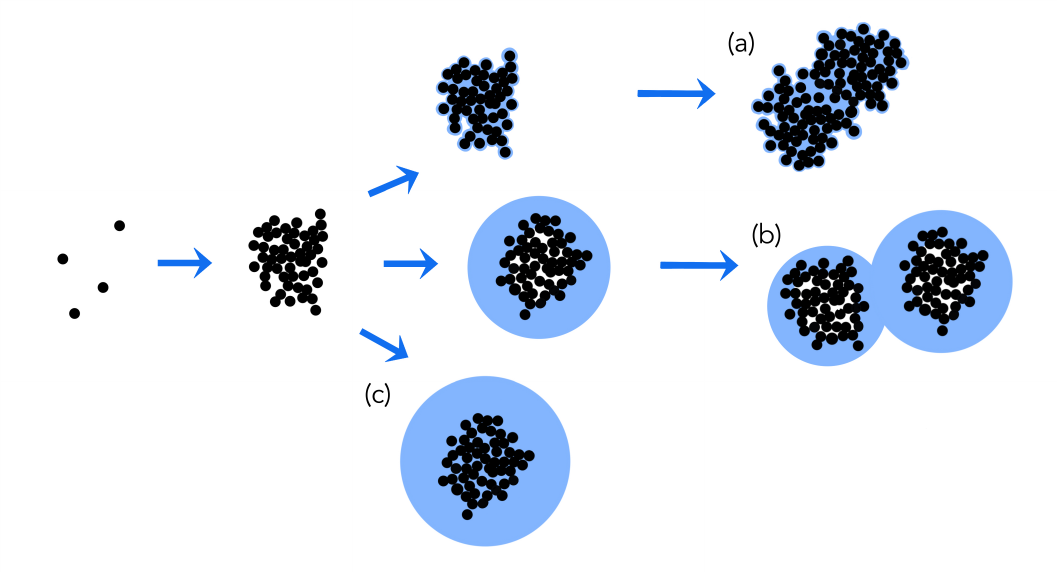}
 \caption{Visual representation of the three ice coating growth scenarios (see Section \ref{sec:materialprops} for a detailed description). In all cases, particles coagulate until they reach a size where ice can form a stable coating on the surface. In the first case (a), the volatile material thinly coats aggregate particles that then undergo growth via coagulation to form larger ice-coated particles with material properties representative of an ice-coated aggregate with small monomer sizes. In the second case (b), the volatile gas is more abundant and able to quickly form particles with a large coating of ice. These particles then coagulate and have fragmentation properties representative of an ice-coated aggregate with large monomer sizes. In the third case (c), a limited number of large particles rapidly form significant ice coatings and grow via condensation. These particles have fragmentation properties representative of a non-aggregate silicate or icy particle}.
 \label{fig:growth_cartoon}
\end{figure*}

Both the binding energy and surface energy can be calculated from either laboratory experiments or numerical models and a range of values exist for these quantities in the literature. In this work, we have conservatively adopted frequently chosen literature values though we note that other values for these parameters are sometimes available and can be different for crystalline versus amorphous molecular structures. In particular, we note that recent laboratory experiments have found that the surface energy of H$_2$O, particularly at temperatures cooler than $\sim$175 K, is significantly lower than previous estimates based on studies performed at higher temperatures \citep{Gundlach2018,Haack2020,Musiolik2021}. However, including these lower surface energy estimates in models of disks, particularly when there is significant turbulence, can inhibit the formation of $\sim$millimeter-sized particles needed to reproduce observed millimeter disk fluxes \citep{Pinilla2020}. This could potentially indicate that the added heat from the kinetic energy of collisions can partially melt the H$_2$O ice mantle, causing an increase in ice stickiness, even at lower temperatures \citep{Nietiadi2020}. Given these uncertainties, future laboratory H$_2$O ice collision experiments at relative velocities and low temperatures relevant for protoplanetary disks, would provide ideal constraints.

CO$_2$ ices also have a range of estimated surface energies from numerical and laboratory work (noting that the value we use in this study, $\gamma_{\rm CO_2}= 60$ erg cm$^{-2}$, has a large uncertainty of $\pm 22$ erg cm$^{-2}$). For example, \citet{Musiolik2016a} find that pure CO$_2$ ice has a critical fragmentation energy comparable to that derived from the tensile strengths of the silicate-containing mixtures basalt and palagonite. However, \citet{Musiolik2016b} find that ice mixtures can have significantly higher surface energies than the value for CO$_2$ used in this work, which can change the collisional behaviour from a homogeneous ice mantle \citep{Musiolik2016a}. Fewer recent constraints exist for the material properties of CO ice, and as such we match and assume the Young's modulus and Poisson's Ratio to that of water as it is in a reasonable range for ice. In general, surface energies in the conditions relevant in protoplanetary disks are not well-constrained as these values are typically based on laboratory experimental setup or idealized numerical computations. 

Furthermore, the structure of ice particles in disks is largely unconstrained. We thus detail below three physically-motivated cases that may occur in protoplanetary disks and lead to significantly different predictions for the size at which particles fragment (see Figure \ref{fig:growth_cartoon}). These cases correspond to particles that have fragmentation properties that resemble those of: (a) an ice-coated aggregate comprised of small monomers, (b) an ice-coated aggregate comprised of large monomers, and (c) an ice-coated compact particle. We further discuss a corollary potential case of an aggregate particle behaving like an ice-free aggregate composed of small monomers. There are several different potential outcomes of ice formation depending on the relative solid to volatile gas abundance, which changes throughout the lifetime of the disk. We assume that either the particles are coated in ice beyond ice lines in the disk, or ice surfaces formed on particles within the molecular cloud from which the star and disk formed and are retained beyond ice lines \citep{Furuya2016,Oberg2021}. Uncertainty remains in how reprocessed these inherited particles from the molecular cloud are, and as such we make no assumption on the composition and structure of particles prior to the disk phase.

In the first potential case (Figure~\ref{fig:growth_cartoon}a), a small coating of ice covers the entirety of the aggregate particle, including the surface of the grain and in the cavities of the aggregate particle. This scenario is likely to occur when there is a sufficient number of particles that can quickly deplete the available supply of volatile gas (or if the gas is only marginally supersaturated) such that each particle is coated with a relatively thin layer of ice. Furthermore, experiments of ice formation on Earth indicate that ice preferentially forms in grain cavities and is thus likely to coat the particle as pictured in Figure~\ref{fig:growth_cartoon}a \citep[e.g.,][]{Campbell2017,Campbell2018,Holden2021}. Once ice formation depletes the supply of volatile material, the abundant large, ice-coated aggregates will undergo further growth via coagulation. This growth is likely to occur with ice-coated aggregates with similar properties \citep[e.g.,][]{Powell2019a}. The resultant large particles that are produced via this growth scenario will then resemble an aggregate particle comprised of small monomers where the connections between the grains are dominated by the material properties of the dominant ice species coating the surface. This case may resemble the structure of a sintered aggregate, however the connection strength is of different compositions (rocky or icy) and the collision outcome between the sintered and unsintered cases will also differ (further discussed in Sections \ref{sec:critical_velocity} and \ref{sec:disc_other_barriers}).

In the second potential case (Figure~\ref{fig:growth_cartoon}b), a large coating of ice will cover the surface of the silicate grains. This scenario is likely to occur if there is an abundant supply of volatile gas than in the first case but there still exists a large number of sufficiently large aggregates that have become ice-coated. In this case, once the particles have depleted the available supply of volatile gas, further particle growth will be dominated by collisions with particles of similar sizes. Due to the large coating of ice on each particle, these collisions will likely result in fragmentation as they are structurally similar to ice-coated aggregates comprised of large monomers (discussed in detail in Section \ref{sec:mon_size_dependnece}). 

In the third potential case (Figure~\ref{fig:growth_cartoon}c), ice will form at a very efficient rate and will quickly dominate the particle's properties. This scenario is likely to occur if volatile gas is abundant and there is a limited supply of large particles. In this case, the limited number of large particles will abundantly form ice and are unlikely to grow further via coagulation due to their low relative number densities. The fragmentation properties of these particles are likely to resemble that of a compact icy grain and not those of an aggregate particle.

In the ice free case, the connections between the monomers are dominated by the material properties of the ice-free monomer cores which are likely composed of grains that are often silicate in composition. This case is likely for small particles and may also occur if ice formation is inefficient or the volatile gas supply is depleted. 

These different structural cases will lead to varying critical fragmentation velocities, ultimately depending on not only the particle composition, but also the particle structure and in particular the size of monomers composing the aggregates (see Figures \ref{fig:monomer_fragmentation_dependance} and \ref{fig:varymonomers} for specifics). For this study, we will evaluate the particle fragmentation for varying compositions, while taking the nominal particle structure to be compact ice-coated unsintered aggregates composed of small $0.1$ $\mu$m sized monomers. This monomer size corresponds to the average particle size expected in the interstellar medium \citep[][]{Blum2006,Gundlach2018}. We take the first case to be our nominal case because large coatings of volatile ice on particles require significant supersaturations of condensible gas which likely only exists in disks at early times or near a species' iceline \citep{Powell2022}. A comparison of how particle structure and monomer size is demonstrated in reference to the critical fragmentation velocities in the following Section (Section \ref{sec:critical_velocity}) with Figures \ref{fig:monomer_fragmentation_dependance} and \ref{fig:composition_limits}.

%%%%%%%%%%%%%%%%%%%%%%%%%%%%%%%%%%%%%%%%%%%%%
%%%%%%%%%%% CRITICAL VELOCITY %%%%%%%%%%%%%%%
%%%%%%%%%%%%%%%%%%%%%%%%%%%%%%%%%%%%%%%%%%%%%

\section{Critical Velocity for Collisional Destruction}\label{sec:critical_velocity}

Particle fragmentation, in the context of the fundamental material properties and structures discussed in Section \ref{sec:materialprops}, occurs in qualitatively different ways for solid rocks versus porous aggregates. For non-aggregate solid rocks with minimal porosity (as in Figure \ref{fig:growth_cartoon}c), cracks typically develop at weak points where atoms are less effectively bound to their neighbors. A sufficiently substantial crack can weaken a solid rock enough for it to fragment into pieces. For porous rocks or aggregates (as in Figure \ref{fig:growth_cartoon}a,b), the connections between distinct, solid monomers are individually weaker than the bonds between component atoms within a monomer, thus fragmentation typically results from breaking connections between monomers. Perhaps counter intuitively, this behavior causes aggregates composed of very small monomers to be more resilient to destructive collisions than solid compact particles since collisions between porous grains tend to compact the aggregate rather than fully breaking it apart along cracks \citep[see][in the case of erosion]{Seizinger2013}. To break an aggregate, the majority of the connections making up the full aggregate must be broken. Sintered aggregates may be stronger to collisions since the connections become fused into a neck as compared to a contact point, but whether the collision will result in growth versus bouncing depends on how compact the aggregate is and where in the disk it is \citep[e.g.,][see Section \ref{discussion} for a discussion of the sintered case]{Maeno1983,Blackford2007,Sirono1999,Sirono2017,Sirono2021}. In protoplanetary disks, particles may begin their growth as fluffy aggregates \citep[e.g.,][]{Smirnov1990,Dominik1997,Blum2000a}, but the size at which growing bodies transition to low-porosity solids remains unclear \citep[][]{Kataoka2013}. We therefore consider both regimes of particle porosity and structure.

For fragmentation of solid non-aggregate particles, we expand on the critical disruption criterion presented in \cite{Stewart2009}. The kinetic energy per mass of a projectile required to destroy a target planetesimal in either the strength or gravity dominated regimes is
\begin{equation}
 Q_{RD}^* = \left(q_s R_{C1}^{9\mu/(3-2\phi)} + q_g R_{C1}^{3\mu}\right) V_i^{2-3\mu}
\end{equation}
The term on the left, $q_s R_{C1}^{9\mu/(3-2\phi)}$, represents the strength regime (small particles bonded together), while the term on right, $q_g R_{C1}^{3\mu}$, represents the gravity regime (rubble piles gravitationally held together). Here $q_s$, $q_g$, $\mu$, and $\phi$ define the material properties in cgs units; values for strong and weak rock are in Table \ref{SL_material_props} \citep[see][for details]{Housen1990,Housen1999,Stewart2009}. The sum of the spherical radii of the target and projectile is $R_{C1}$, which we approximate as the size of the larger colliding particle. The disruption criterion for our particles is not expected to be dominated by the gravity regime, however it is still included in the model. 

\begin{deluxetable}{lllll} [!htb]
\tablecolumns{5}
\tablecaption{Fragmentation Criterion Material Properties \citep{Stewart2009} \label{SL_material_props}}
\tablehead{ % column headings
 \colhead{Rock Type} &
 \colhead{$q_s$} &
 \colhead{$q_g$} &
 \colhead{$\mu$} &
 \colhead{$\phi$}
}
\startdata
Strong Rock & $7 \times 10^4$ & $10^{-4}$ & 0.5 & 8 \\
Weak Rock & 500 & $10^{-4}$ & 0.4 & 7 \\
\enddata 
\end{deluxetable}

The disruption criterion can be described as the binding energy keeping together the compact particle (or in the gravity regime, a porous rubble pile consisting of a gravitational aggregate of rigid spheres) \citep{Stewart2009}. Comparing this energy with the collisional kinetic energy of a projectile provides the critical velocity needed to make cracks in and ultimately break apart the target particle. Here the kinetic energy of the projectile is $KE = (1/2)m_{\mathrm{p}}v_{\mathrm{rel}}^2$, while the binding energy of the target particle is $BE = m_{\mathrm{t}}Q_{RD}^*$, where $m_{\mathrm{p}}$ and $m_{\mathrm{t}}$ are the masses of the projectile and target respectively. Converting the masses in terms of the target radius, $R_{C1}$, and solving for the relative velocity $v_{\mathrm{rel}}$, we find that the critical relative velocity required to break apart a solid compact particle is
\begin{equation} \label{eq:vcrit_SL}
 v_{\mathrm{crit,SL}} = \sqrt{\frac{2}{f^3} Q_{RD}^*}
\end{equation}
where $f$ is the size ratio between the two colliding particles. It should be mentioned that for solid compact particles, relative velocities that are nearly at critical fragmentation velocities may also result in bouncing rather than sticking, which is further discussed in Section \ref{discussion}.

For fragmentation of small aggregate bodies (not gravitational rubble piles), we follow the work of \cite{Wada2007,Wada2009b} to find the critical velocity for collisional destruction. Their numerical studies model the collisions between aggregates in the ballistic cluster--cluster aggregation (BCCA) and ballistic particle--cluster aggregation (BPCA) regimes, or more generally fluffy and compact aggregates. This kind of fragmentation focuses on pulling apart each monomer, in contrast to the solid fragmentation framework of \cite{Stewart2009}. The energy required to break apart two contact monomers is 
\begin{equation}
 E_{\text{break}} = 1.54F_{\text{c}}\delta_{\text{c}} \;,
\end{equation}
where $F_{\text{c}} = 3\pi \gamma R$ is the maximum force required to separate the contact and $\delta_{\text{c}} = (9/16)^{1/3}a_0^2/(3R)$ is the critical separation or compression distance between the two monomers (denoted with numerical subscripts). These expressions depend on the contact circle radius $a_0 = (9\pi\gamma R/E^*)^{2/3}$ \citep{Wada2007}, $1/R=1/r_{\rm m,1} + 1/r_{\rm m,2}$ with $r_{\rm m,1}$ and $r_{\rm m,2}$ being the monomer radii, and $1/E^*=(1-\nu_1^2)/E_1 + (1-\nu_2^2)/E_2$ \citep{Wada2009b}. All of these are a function of the material properties---Young's modulus ($E$), Poisson's ratio ($\nu$), surface energy ($\gamma$), and the material density ($\rho_i$) discussed in Section \ref{sec:materialprops}. The kinetic energy of a colliding aggregate is $KE = (1/2)N_{\text{total}}m_{\rm m}(v_{\text{rel}}/2)^2$, where $N_{\text{total}}$ is the total number of particles composing both aggregates and $m_{\rm m} = (4/3)\pi r_{\rm m}^3\rho_i$ is the mass of a monomer. The critical impact energy is $E_{\text{crit}} = kN_{\text{total}}E_{\text{break}}$, where $k$ is a dimensionless factor which takes into account whether the aggregate is fluffy or compact. \cite{Wada2009b} find that for fluffy aggregates (BCCA) $k \sim 10$, while for compact aggregates (BPCA) $k \sim 30$. The critical relative velocity is then found by balancing the kinetic energy with the critical impact energy and solving.
\begin{equation}\label{eq:vcrit_wada}
 v_{\text{crit,W}} = \sqrt{\frac{8kE_{\text{break}}}{m_{\rm m}}}
\end{equation}

Rolling friction is not included in our model as it may not impact the outcome of collisions as demonstrated by \cite{Arakawa2022}. We also note that viscous energy dissipation through the particle may not necessarily break apart every monomer connection in an aggregate at this particular critical fragmentation velocity as other tangential forces and impact parameter may determine the collision outcome depending on the assumed viscous dissipation timescale \citep[e.g.,][]{Arakawa2022a}. This framework does however provide an intuition for the compositional dependencies on fragmentation.

\begin{figure}[!htb]
 \centering
 \includegraphics[width=8.5cm]{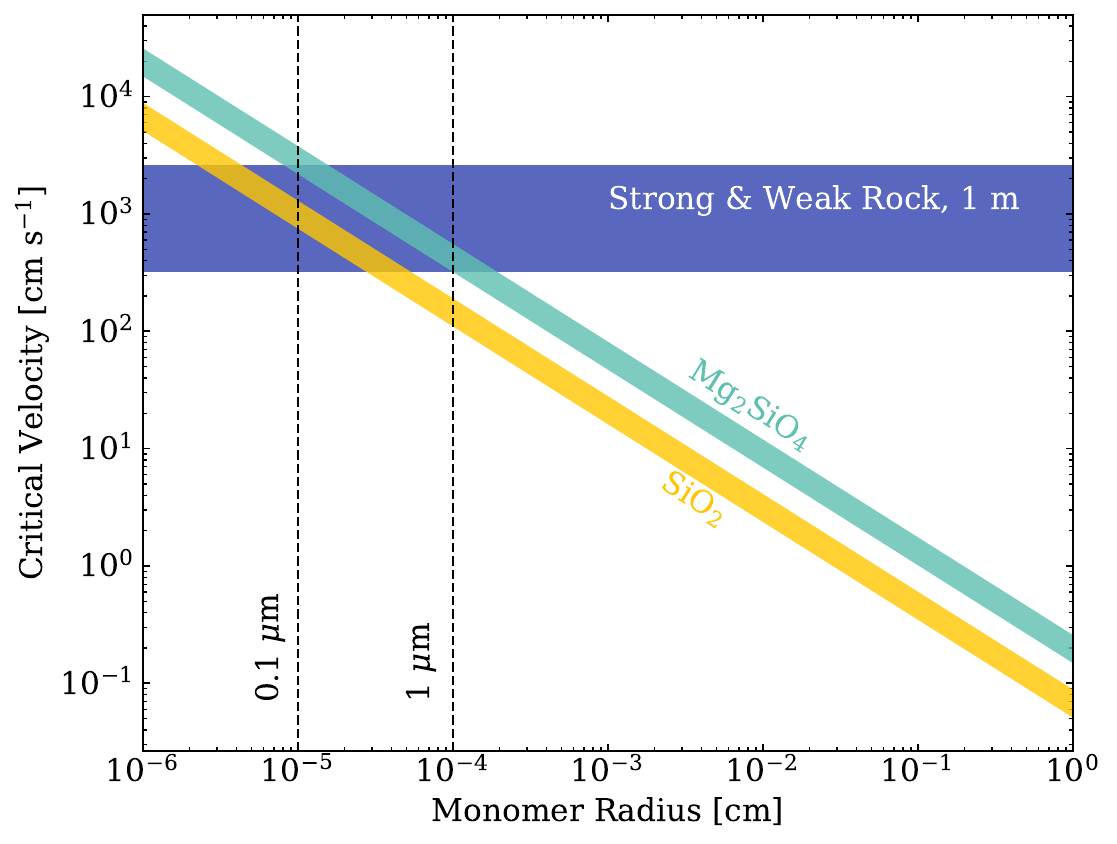}
 \caption{The critical velocity of aggregates for all compositions strongly depends on the monomer size. The SiO$_2$ and Mg$_2$SiO$_4$ regions are the critical fragmentation velocities for silicate aggregate particles following \cite{Wada2007,Wada2009b} with the lower bound and upper bound at each monomer size defined by BCCA and BPCA clusters respectively. The Strong \& Weak Rock region is the critical fragmentation velocity for a solid non-aggregate meter-sized rock following the \cite{Stewart2009} energy calculations using values from Table \ref{SL_material_props}. }
 \label{fig:monomer_fragmentation_dependance}
\end{figure}

Figure \ref{fig:monomer_fragmentation_dependance} compares the calculated critical velocities for non-aggregate and aggregate particles, highlighting the impact of monomer size. The critical velocities for aggregate silicates composed of SiO$_2$ or Mg$_2$SiO$_4$, with monomer sizes in the range of roughly 0.1 to 1 microns, are comparable to those of non-aggregate strong and weak rock as demonstrated in Figure \ref{fig:monomer_fragmentation_dependance}. As the monomer size increases the critical velocity for aggregates monotonically decreases, indicating that the choice of monomer size is an important parameter in modelling fragmentation (results discussed in \ref{sec:mon_size_dependnece}). As mentioned at the end of Section \ref{sec:materialprops}, we choose a fiducial monomer size of 0.1 $\mu$m, consistent with particles in the ISM \citep{Oberg2021}.

\begin{figure}[!htb]
 \centering
 \includegraphics[width=8.5cm]{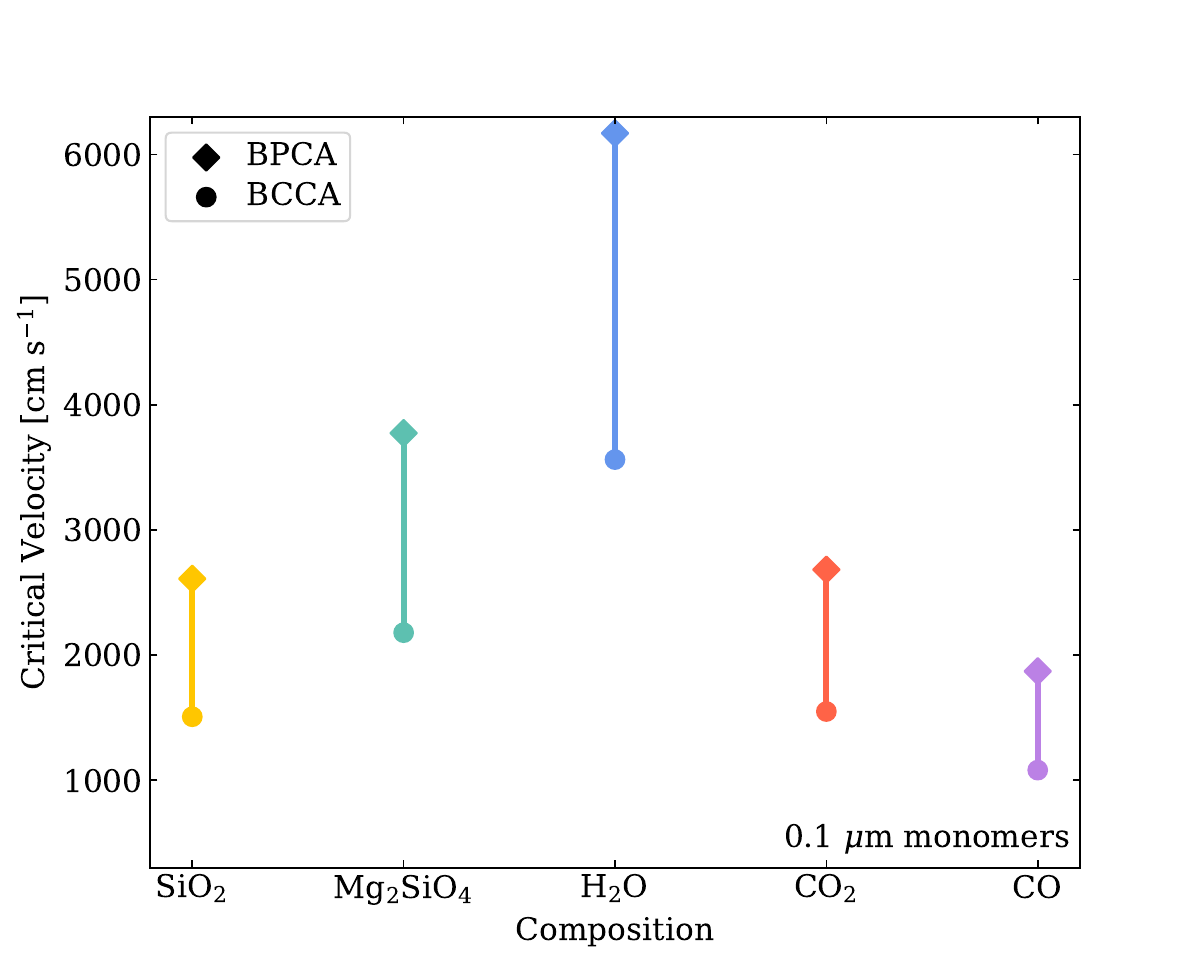}
 \caption{Different aggregate compositions define the overall strength of particles. Compact BPCA H$_2$O ice is the strongest aggregate. Because strength is determined by surface interactions between monomers, silicates coated in ice will have strengths determined by that ice.}
 \label{fig:composition_limits}
\end{figure}

The difference in surface energy is also demonstrated as the critical velocities are consistently lower for SiO$_2$ than Mg$_2$SiO$_4$, where SiO$_2$ has a much lower surface energy (see Table \ref{material_props}).  Figure \ref{fig:composition_limits} compares the differing critical fragmentation velocities between fluffy BCCA and compact BPCA aggregate particles with 0.1$\mu$m monomers and varying compositions.

%%%%%%%%%%%%%%%%%%%%%%%%%%%%%%%%%%%%%%%%%%%%%
%%%%%%%%%%%%%% DISK PROPERTIES %%%%%%%%%%%%%%
%%%%%%%%%%%%%%%%%%%%%%%%%%%%%%%%%%%%%%%%%%%%%

\section{Protoplanetary Disk Model Properties} \label{sec:disk_params}

\subsection{Surface Density and Temperature Profile} \label{ssec:surf. density and temp}
In our modeling, we use the surface density for TW Hya from \cite{Powell2019a}, in comparison with the MMSN, to demonstrate the importance of disk properties in shaping particle evolution. The disk surface density profile for TW Hya comes from the \cite{Lynden-Bell1974} and \cite{Hartmann1998} self-similar solution to the viscous equations:
\begin{equation} \label{eq:sigma TW Hya}
 \Sigma = \Sigma_o\left(\frac{r}{r_{\mathrm{c}}}\right)^{-\gamma}\exp\left[{-\left(\frac{r}{r_{\mathrm{c}}}\right)^{2-\gamma}}\right]
\end{equation}
where $\Sigma_o = 175$ g cm$^{-2}$ is the derived surface density constant, $r_{\mathrm{c}} = 30$ AU is the critical radius, and $\gamma = 1$ is the power law parameter. The surface density profile for the MMSN follows the standard prescription from \cite{Weidenschilling1977} and \cite{Hayashi1981}:
\begin{equation}\label{eq:sigma MMSN}
  \Sigma_{\mathrm{MMSN}} = 1700\: \mathrm{g}\: \mathrm{cm}^2\: \left(\frac{r}{\mathrm{1 AU}}\right)^{-3/2}
\end{equation}

Protoplanetary disks are defined as either passively or actively heated depending on whether accretion onto the host-star is present. In a passively heated disk, stellar irradiation is the dominant heat source for the entire disk, where incoming radiation from the host star is absorbed in local regions of the disk and then re-emitted as a blackbody. An actively heated disk includes viscous heating caused by viscous midplane accretion of the disk onto the host star. If viscous accretion heating is included, it will be the dominant heat source in the inner disk while the mid to outer disk remains dominated by irradiation heating. Throughout a disk's lifetime, the disk can either remain fully passive, fully active, or switch from one to the other \citep{Armitage2017a}. 

From \cite{Chiang1997}, the irradiation temperature follows a power-law depending on the stellar mass ($M_*$) and luminosity ($L_*$)
\begin{equation} \label{eq.T_irad}
 T_{\text{irradiation}} = T_o \left(\frac{r}{\mathrm{1 AU}}\right)^{-3/7} \: ,
\end{equation}
\begin{equation}
 T_o = \bigg(\frac{2}{7}\bigg)^{1/4} \bigg(\frac{L_*}{4 \pi\sigma_{\mathrm{SB}}}\bigg)^{2/7} \bigg(\frac{k}{\mu GM_*}\bigg)^{1/7}
\end{equation}
where $\sigma_{\mathrm{SB}}$ is the Stefan-Boltzmann constant, $k$ is the Boltzmann constant, $G$ is the gravitational constant, $\mu = 2.3m_{\mathrm{H}}$ is the mean molecular weight assuming a hydrogen-helium composition, and $m_{\rm H}$ is the mass of a hydrogen atom. Using values from \cite{Powell2019a}, TW Hya has $M_* = 0.8 M_{\odot}$, $L_* = 0.28L_{\odot}$, and $T_o = 82$ K. The MMSN has $T_o = 120$ K \cite{Chiang1997}.

The temperature due to viscous heating is determined by the disk's vertical optical depth ($\tau_{\text{vert}}$) and gas surface density ($\Sigma$, Eq. \ref{eq:sigma TW Hya},\ref{eq:sigma MMSN})
\begin{equation} \label{eq.T_acc}
 T_{\text{accretion}} = \bigg[ \frac{9}{32\pi}\frac{\tau_{\text{vert}}}{\sigma_{\mathrm{SB}}}\dot{M}\Omega^2 \bigg]^{1/4} \;,
\end{equation}
\begin{equation}
    \tau_{\text{vert}}=\frac{1}{2} \Sigma \kappa
\end{equation}
where $\kappa = 0.5$ cm$^2$ g$^{-1}$ is the opacity, $\dot{M} = 10^{-8}$ M$_\odot$ yr$^{-1}$ is the standard observed mass accretion rate, and $\Omega=\sqrt{GM_*/r^3}$ is the orbital angular frequency \citep[e.g.,][]{Garaud2007,Kratter2010,Oka2011,Kratter2011,Rosenthal2020}. 

The total temperature including accretion is then 
\begin{equation} \label{temp eq}
 T = (T_{\text{irradiation}}^4 + T_{\text{accretion}}^4)^{1/4}
\end{equation}
where for a passive disk $T_{accretion}$ is set to 0.

The midplane temperature profile is a key component of the model as all governing processes involve the temperature. In particular, the temperature can significantly alter ice line locations as viscous accretion heating pushes ice lines further out in the disk, especially for species with higher sublimation temperatures.

\begin{figure}[!htb]
 \centering
 \includegraphics[width=8.5cm]{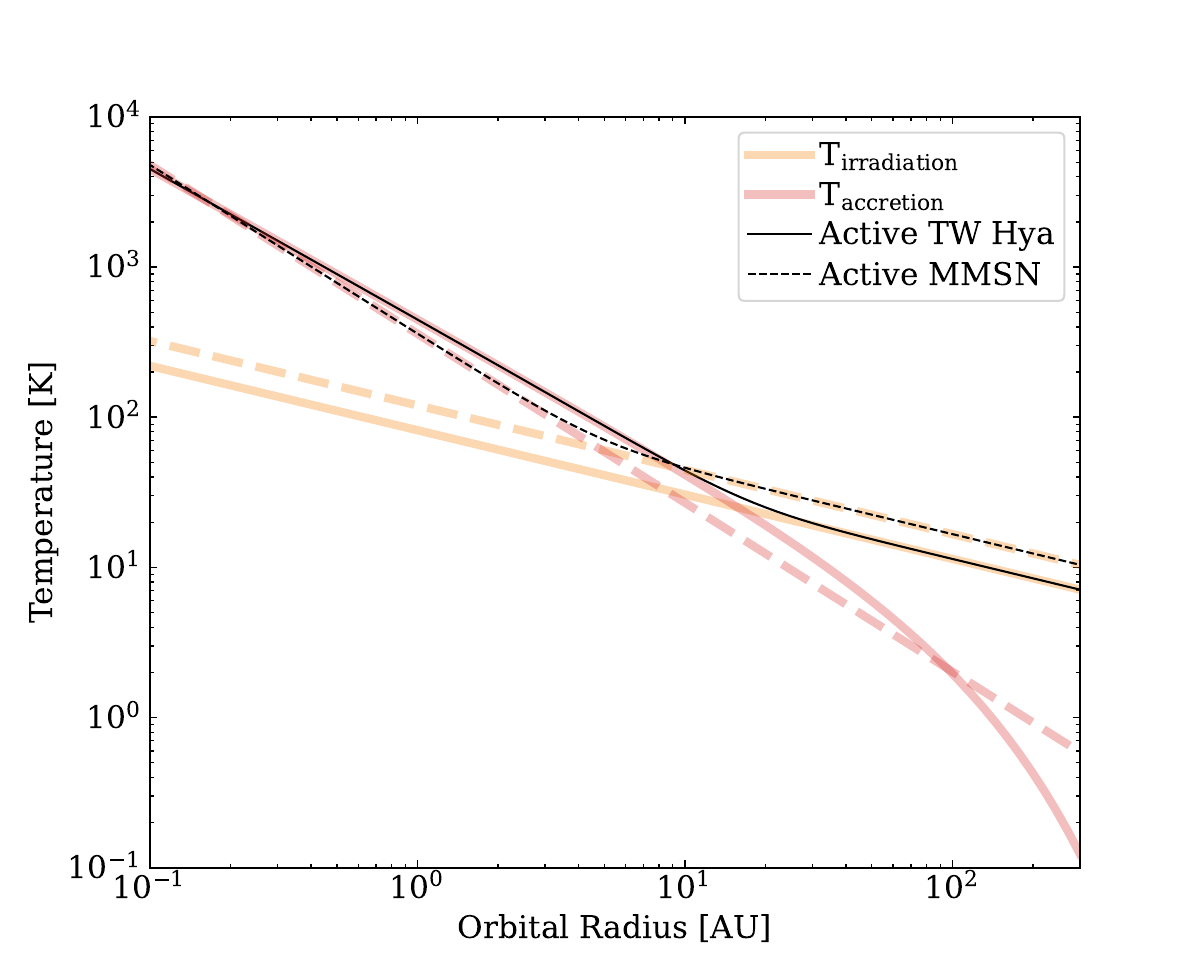}
 \caption{For an active disk, the inner disk follows the viscous accretion heating profile (red) while the outer disk follows the passive stellar irradiation heating profile (orange). The temperature profiles for both the MMSN and TW Hya are displayed with dashed and solid lines respectively.}
 \label{fig:temp_prof}
\end{figure}

\subsection{Ice lines}\label{subsec: icelines}

Ice lines are the radial and vertical locations in the disk where molecules freeze out onto silicate particles. We calculate the radial ice line locations for H$_2$O, CO$_2$, and CO, which are abundant volatiles in protoplanetary disks (see \citealt{Oberg2021} for a review of disk chemistry). 

An ice line is determined by the temperature (and to some extent disk surface density) at which a molecule's adsorption and desorption fluxes are balanced \citep[see][]{ Hollenbach2008, Oberg2011, Powell2017, Oberg2019}. This temperature is compared to the disk temperature profile in order to find the specific radius at which the fluxes are in steady state. The expressions for the fluxes (number of molecules per area per time) are:
\begin{equation}
 F_{\mathrm{adsorp}} \sim n_{\mathrm{i}}c_{\mathrm{s}}
\end{equation}
\begin{equation}
 F_{\mathrm{desorp}} \sim N_{\mathrm{s,i}}\nu_{\mathrm{vib}}\text{e}^{E_{\mathrm{i}}/kT_{\mathrm{grain}}}f_{\mathrm{s,i}}
\end{equation}
where $n_{\mathrm{i}}$ is the molecular gas number density and $c_{\mathrm{s}} = \sqrt{kT/\mu}$ is the isothermal sound speed. The number of adsorption sites per cm$^2$ on the particle is $N_{\mathrm{s,i}}=10^{15}$ sites per cm$^2$. The vibrational frequency of the molecules in the surface potential well is $\nu_{\mathrm{vib}} = 1.6 \times 10^{11}\sqrt{E_{\mathrm{i}}/\mu_{\mathrm{i}}}$ s$^{-1}$, with $E_{\mathrm{i}}$ being the adsorption binding energy of the molecule in units of Kelvin and $\mu_{\mathrm{i}}$ being the molecular weight in grams. We take the fraction of occupied adsorption sites, $f_{\mathrm{s,i}}$, to be unity. The grain temperature, $T_{\mathrm{grain}}$, is assumed to be the same as the midplane temperature (see Eq.~(\ref{temp eq})). The adsorption and desorption fluxes depend strongly on the species molecular properties, which are listed in Table~\ref{table: ice line_params}. Ice lines are key in determining which species are present throughout the disk.

%%%%%%%%%%%%%%%%%%%%%%%%%%%%%%%%%%%%%%%%%%%%%
%%%%%%%%%%%% RELATIVE VELOCITY %%%%%%%%%%%%%%
%%%%%%%%%%%%%%%%%%%%%%%%%%%%%%%%%%%%%%%%%%%%%

\section{Relative Velocities of Solid Particles in Protoplanetary Disks}\label{sec:relative_velocity}

The motion of small particles in a protoplanetary disk is primarily determined by the motion of the surrounding gas and by how strongly the particles move with the gas via gas drag. We calculate particle--particle relative velocities, following the works of \citet{Whipple1972}, \citet{Weidenschilling1977a}, \citet{Chiang2010}, and \citet{Perets2011}. Particles throughout protoplanetary disks can be governed by three different drag regimes: Epstein, Stokes, and ram pressure. When the particle radius, $s$, is less than the mean free path, $\lambda$, of the gas so that $s<(9/4)\lambda$, the particle is in the Epstein drag regime. Typically, Epstein drag applies to particles in the outer disk that are well-coupled to the gas. For larger particles with $s\ge (9/4)\lambda$, the governing drag regime must be determined using the Reynolds number, $Re=2s v_{\rm rel}/\nu$, where the kinematic viscosity $\nu = 0.5\lambda \bar v_{\rm th}$, and $\bar{v}_{\text{th}} = \sqrt{8/\pi}c_{\mathrm{s}}$ is the thermal velocity of the gas. When $Re < 1$ the particle is in the Stokes drag regime, while particles with $Re > 800$ are in the ram pressure drag regime. We approximate particles in the intermediate Reynolds number range by linearly extending the Stokes and ram pressure regimes and find the transition between the two to be where the two analytic coefficient of drag expressions intersect, $Re\sim54$. At this intersection, the approximation deviates from the intermediate regime by a factor of $\sim4$ (see Appendix~\ref{re derivation} for details). 

The drag forces in each regime are given by:
\begin{equation} \label{eq:f_d}
 F_{\text{d}} = \left\{
\begin{array}{ll}
 {\frac{4}{3}\pi\rho_{\text{g}} s^2 \bar{v}_{\text{th}} v} & \text{Epstein} \\ 
 {3\pi\rho_{\text{g}} s \lambda \bar{v}_{\text{th}} v} & \text{Stokes} \\
 {0.22\pi\rho_{\text{g}} s^2 v^2} & \text{Ram pressure}
\end{array} 
\right.
\end{equation}
where $\rho_{\text{g}}$ is the density of the gas and $v$ is the relative velocity between the particle and the gas. 

A particle is considered well-coupled to the surrounding gas if its orbital period is longer than the time it takes gas drag to stop the particle's motion. This is called the stopping time, and is found by dividing the particle momentum by the drag force.
\begin{equation} \label{eq:t_s}
 t_{\text{s}} = \frac{mv}{F_{\mathrm{d}}} \approx \left\{
\begin{array}{ll}
 {\left(\frac{\rho_{\text{s}}}{\rho_{\text{g}}}\right) \frac{s}{\bar{v}_{\text{th}}}} & \text{Epstein} \\ 
 {\frac{4}{9}\left(\frac{\rho_{\text{s}}}{\rho_{\text{g}}}\right) \frac{s^2}{\lambda \bar{v}_{\text{th}}}} & \text{Stokes} \\
 {\left(\frac{\rho_{\text{s}}}{\rho_{\text{g}}}\right) \frac{s}{v}} & \text{Ram}
\end{array} 
\right. 
\end{equation}
The dimensionless stopping time $\tau \equiv t_{\text{s}}\Omega$ is used throughout the model, and is also known as the Stokes number. Well-coupled particles have $\tau < 1$. The density of the particle is $\rho_{\rm s} = \phi_{\rm s}\rho_{i}$, where $\phi_{\rm s}$ is the particle filling factor and $\rho_{i}$ is the material density (see Appendix \ref{A: size ratio and porosity}). 

\begin{figure}[!htb]
 \centering
 \includegraphics[width=8.5cm]{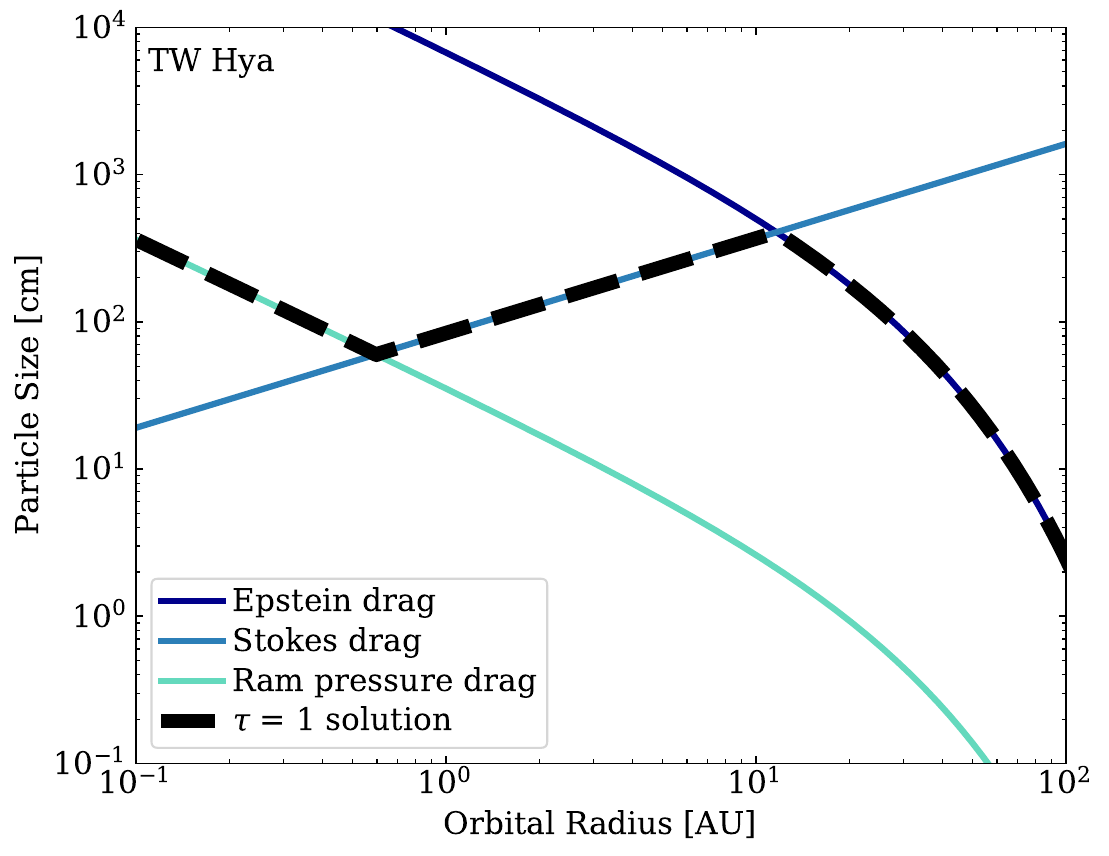}
 \includegraphics[width=8.4cm]{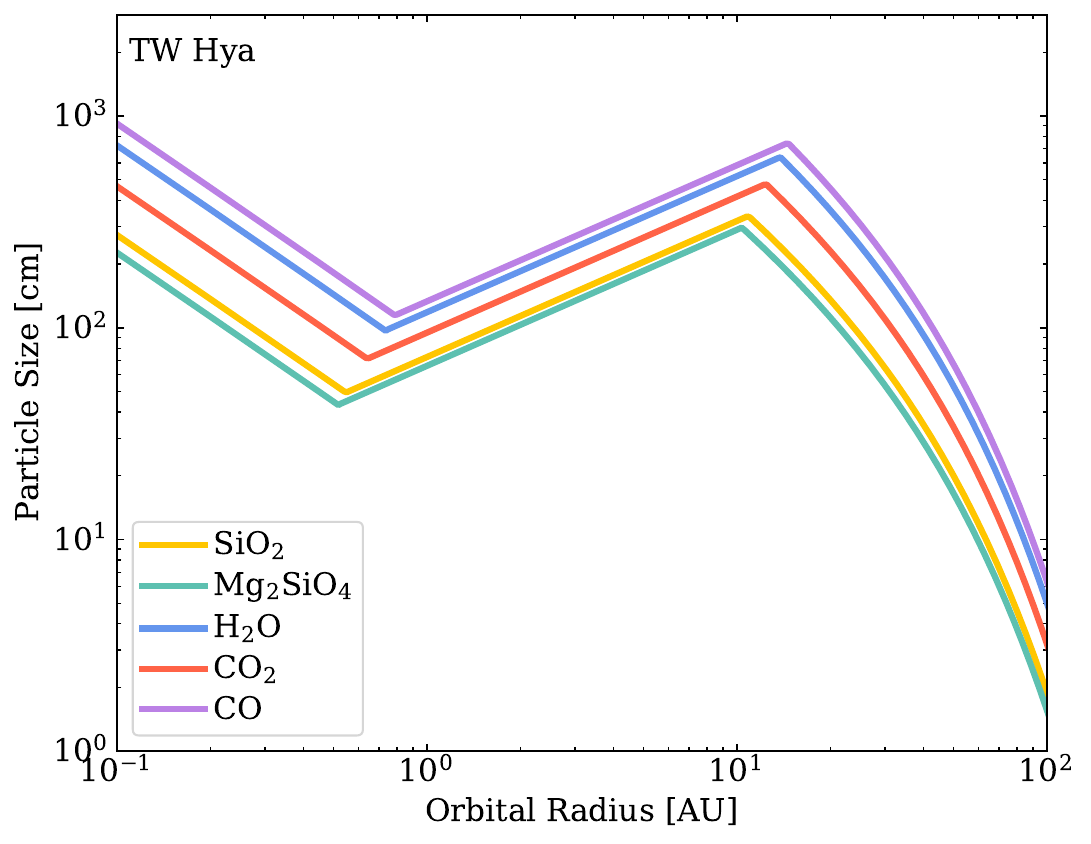} 
 \caption{Left: Comparison of the Epstein, Stokes, and ram pressure drag regimes for $\tau=1$ particles. The dashed line represents the $\tau=1$ solution and serves as an intuitive proxy for the fragmentation region dependencies on drag regime. Right: The $\tau=1$ solution varies with composition due to different material densities assuming $\phi_{\rm s}=0.3$}.
 \label{fig:TWHya_tau1}
\end{figure}

Marginally coupled particles, approximated by $\tau=1$, can be examined to build intuition for how the different drag regimes affect the sizes at which growing and drifting particles fragment due to collisions. These particles move at the highest velocities relative to the gas and are thus most likely to experience high energy  particle--particle collisions that lead to fragmentation. The particle size at which $\tau=1$ as a function of orbital radius (Figure~\ref{fig:TWHya_tau1}) can be found by solving for the sizes in Equation~(\ref{eq:t_s}) and applying the drag conditions discussed above. Figure~\ref{fig:TWHya_tau1} compares the $\tau=1$ particles for each drag regime, displays the drag-dependant $\tau=1$ solution, and demonstrates how different particle compositions change the solution. A particle in the inner disk is dominated by the ram pressure regime, the central disk is dominated by the Stokes regime, and the outer disk is dominated by the Epstein regime. This simple analytic framework provides some intuition for where in the disk and at which particle sizes fragmentation may occur.

\begin{figure*}[!htb]
 \centering
 \includegraphics[width=17cm]{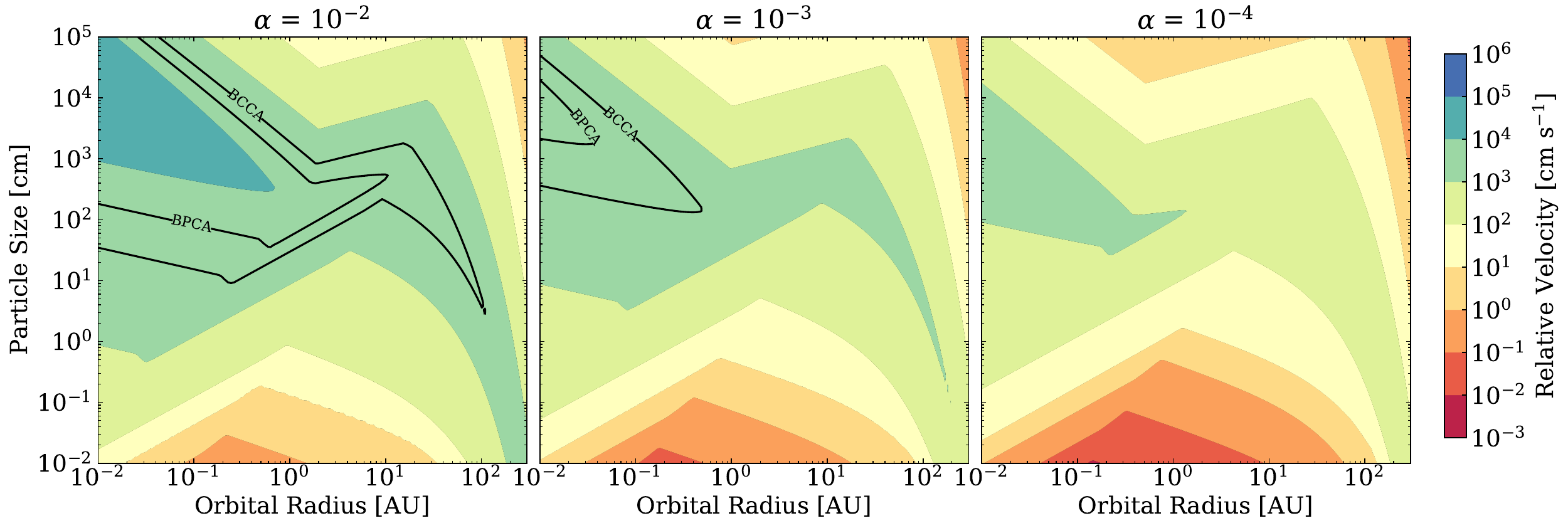}
 \includegraphics[width=17cm]{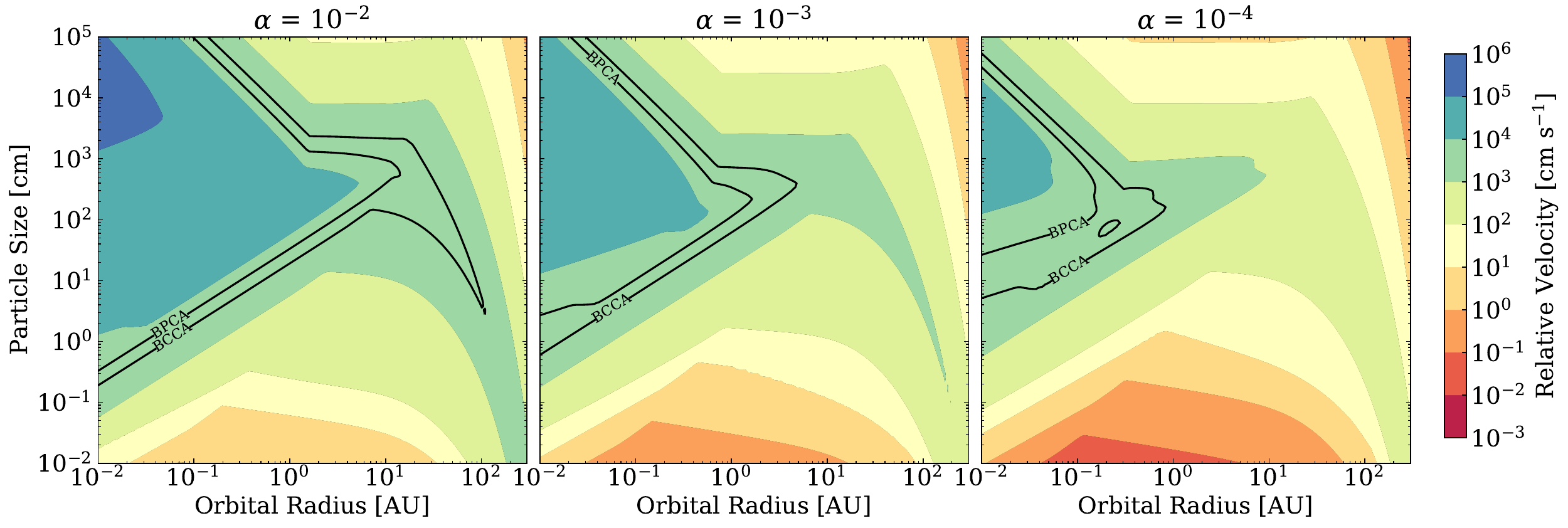} 
 \caption{Relative velocities are higher, and fragmentation is easier, for actively-heated disks with stronger turbulence.  particle--particle relative velocities are displayed as a function of orbital radius and particle size for a passive (top) and active (bottom) disk. The columns correspond to varying turbulence parameterized by an $\alpha$ of $10^{-2}$, $10^{-3}$, and $10^{-4}$ from left to right. Critical fragmentation velocities for BCCA and BPCA H$_2$O solid aggregate particles are displayed as black labelled lines. Varying profiles are shaped by transitions in drag regimes (see Figure~\ref{fig:TWHya_tau1}). Particle size refers to the radius of the target particle, which in this case collides with a projectile half its size.}
 \label{fig:vrels_alpha}
\end{figure*}

\subsection{Particle--Particle Relative Velocities}\label{subsec:vrel_deriv}

Particle--particle relative velocities depend on particle size ($s$) and orbital radius ($r$) via the dimensionless stopping time ($\tau$). We apply analytic relative velocity expressions which are valid for all particle sizes and drag regimes throughout the disk. We numerically compute the stopping time for each particle through an iterative process which uses the particle's size and position in the disk to determine the relevant drag force governing its motion.

There are two components to the relative velocity between two particles ($v_{\rm rel}$): the relative laminar drift velocity and the relative velocity that arises due to turbulent motion of gas in the disk. The total relative velocity is the vector summation of these two components.
\begin{equation} \label{eq.10}
 v_{\text{rel}} = \sqrt{v_{\text{laminar}}(\tau)^2 + v_{\text{turbulent}}(\tau,\alpha)^2}
\end{equation}

Laminar drift arises in disks as the gas orbits at sub-Keplerian velocities due to a radial pressure gradient. Solid particles which would otherwise orbit at the Keplerian velocity feel a headwind due to gas drag causing the particles to lose angular momentum and drift inward. The laminar drift velocity of an individual particle includes both radial ($v_r$) and azimuthal ($v_{\phi}$) disk components which incorporate the inward drift and the orbital velocity \citep[for a review, see][]{Chiang2010}:
\begin{equation}\label{eq.vr}
 v_r = -2\eta v_\mathrm{K}\left(\frac{\tau}{1+\tau^2}\right)
\end{equation}
\begin{equation}\label{eq.vphi}
 v_\phi = -\eta v_\mathrm{K} \left(\frac{1}{1+\tau^2}\right) \;\;,
\end{equation} 
where $\eta = 0.5 c_s^2 / v_{\mathrm{K}}^2 = 0.5 H/r$ is the gas-pressure support parameter ($H=c_s/\Omega$ is the disk gas scale height) and $v_{\mathrm{K}} = \sqrt{GM_*/r}$ is the Keplerian velocity \citep{Weidenschilling1977a}. With these components, the relative laminar drift velocity between two particles is then
\begin{equation} \label{eq:vlam}
 v_\mathrm{laminar} = \sqrt{(v_{r,1}-v_{r,2})^2 + (v_{\phi,1}-v_{\phi,2})^2} \;,
\end{equation}
where the subscripts 1 and 2 refer to evaluation of Equations (\ref{eq.vr}) and (\ref{eq.vphi}) for particle 1 and 2, respectively.

We compute the relative velocity due to particle interactions with turbulent gas following the framework presented in \cite{Ormel2007}, specifically Equations (16) and (21d) (see Appendix A of \cite{Powell2019a} for details). When using Equation (16), we assume that the overturn time of the largest eddy is $t_L=1$ and separately calculate the various times based on the dimensionless stopping time of both colliding particles. For typical particle size distributions, collisions of comparably-sized particles both lead to the greatest rate of particle growth and the highest likelihood of collisional disruption. The gas velocity is taken to be $v_{\rm gas}=\sqrt{\alpha} \bar{v}_{\rm th}$, where $\alpha$ is the \cite{Shakura1973} accretion disk turbulence parameter. Expected protoplanetary disk values for $\alpha$ lie in the range of $10^{-5}$ to $10^{-2}$ \citep{Andrews2020}. Note that in our model, the $\alpha$ that parametrizes turbulent motion is a free parameter. In particular, given uncertainties in disk accretion models, we do not require $\alpha$ to have the same value that would be needed to model viscous accretion and we separately choose $\dot M$ in Equation~(\ref{eq.T_acc}). For reference, using the values given for TW Hya in Section \ref{sec:disk_params} evaluated at $r = 3$AU, the accretion equation $\dot M = 3\pi \Sigma \nu_t$, with turbulent viscosity $\nu_t = \alpha c_s H$, implies $\alpha = 3\times 10^{-4}$. 
The impact of turbulence on particle--particle relative velocity is displayed for passively and actively heated disks in Figure~\ref{fig:vrels_alpha} with $\alpha$'s of $10^{-2}$, $10^{-3}$, and $10^{-4}$. Ultimately, the larger the turbulence, the larger the relative velocity, and with larger relative velocities fragmentation is more likely. For the remainder of this work, we choose a fiducial value of $\alpha=10^{-3}$.

We note that particle settling may result in a minimum level of gas turbulence and hence a minimum expected $\alpha$. Though we do not explicitly enforce this limit in our model, we comment on its magnitude here. Turbulence due to Kelvin-Helmholtz shear instability limits particle settling to the midplane of protoplanetary disks \citep[e.g.,][]{Weidenschilling1980,Sekiya1998, Sekiya2000, Sekiya2001,Youdin2002,Chiang2008}. The instability arises when the Richardson number \citep{Chandrasekhar1961} drops to a critical value of $\mathrm{Ri}\sim1$ \citep{Johansen2006a}, corresponding to a maximum particle scale height (for a large midplane dust-to-gas ratio) of $H_{\rm p} \approx \eta \mathrm{Ri}^{1/2} r \sim \eta r$ \citep{Chiang2008,Gerbig2020}. In our model, the particle scale height is calculated using \citep{Ormel2012}
\begin{equation}
  H_{\rm p} = H\sqrt{\frac{\alpha}{\alpha + \tau}}  \;\;.
\end{equation}
Setting $H_{\rm p}$ to its minimum allowed value yields $\alpha \sim 0.5\eta\tau/(1-\eta)$. To drive the shear instability, particles must be sufficiently coupled to the gas to affect gas motion, so this effect generates maximum turbulence when particles have $\tau \sim 1$ and (since $\eta \ll 1$), $\alpha \sim 0.5\eta$. As an example: at $r = 3$AU for the TW Hya parameters used above, $\eta = 4\times 10^{-4}$ for a passively heated disk and $\eta = 10^{-3}$ for an actively heated disk with $\dot M = 10^{-8} M_\odot/$yr.

Table \ref{tab:fiduc_params} includes a list of the fiducial parameters and their values used in calculating relative velocities and corresponding fragmentation regions in Figures \ref{fig:vrels_alpha}, \ref{fig:frag_comp}, \ref{fig:frag_SL}, \ref{fig:frag_explain}, and \ref{fig:frag_all}. See Appendix \ref{A: size ratio and porosity} for a discussion of the effects of varying size ratio ($\chi$) and filling factor ($\phi_{\rm s}$) on the relative velocity.

\begin{deluxetable}{lll} [!htb]
\tablecolumns{3}
\tablecaption{Fiducial model parameters \label{tab:fiduc_params}}
\tablehead{ 
 \colhead{Description} &
 \colhead{Symbol} &
 \colhead{Value} 
}
\startdata
dust-to-gas ratio & $f_{\rm d}$ & $10^{-2}$ (early disk times)\\ 
& & $10^{-3}$ (later disk times)  \\
turbulence parameter & $\alpha$ & $10^{-3}$ \\
particle size ratio & $\chi$ & 0.5 \\
particle filling factor & $\phi_{\rm s}$ & 0.3\\
monomer size & $r_{\rm m}$ & 0.1 $\mu$m
\enddata 
\end{deluxetable}

%%%%%%%%%%%%%%%%%%%%%%%%%%%%%%%%%%%%%%%%%%%%%
%%%%%%%%%%%%%%% FRAG REGIONS %%%%%%%%%%%%%%%%
%%%%%%%%%%%%%%%%%%%%%%%%%%%%%%%%%%%%%%%%%%%%%

\section{Species Dependent Fragmentation Regions} \label{sec:frag regions}

Particles undergo collisional fragmentation if their relative velocity ($v_{\mathrm{rel}}$, derived in Section \ref{sec:relative_velocity}) in the disk reaches or exceeds the species- and structure-dependant critical fragmentation velocity ($v_{\mathrm{crit}}$, derived in Section \ref{sec:critical_velocity}). By defining fragmentation as where these two velocities are the same, we can solve for the sizes at which collisions will result in fragmentation at each orbital radius. In the inner disk, there exist two fragmentation sizes with $v_{\mathrm{rel}} = v_{\mathrm{crit}}$ for each orbital radius, bounding a fragmentation region where $v_{\mathrm{rel}} \geq v_{\mathrm{crit}}$ centered on particles with $\tau =1$. Farther out in the disk, the range of particle sizes that fragment becomes smaller. For some material and disk properties, fragmentation ceases altogether in the outer disk. Using this framework, we can create regions of fragmentation in TW Hya for aggregate particles (based on Equation \ref{eq:vcrit_wada}) composed of SiO$_2$, Mg$_2$SiO$_4$, H$_2$O, CO$_2$, and CO, as well as for strong and weak non-aggregate solid particles (based on Equation \ref{eq:vcrit_SL}), shown in Figures \ref{fig:frag_comp} and \ref{fig:frag_SL} respectively. The fragmentation regions are recreated for the MMSN in Appendix \ref{A: MMSN} for comparison. We note that even if particles do not reach critical fragmentation velocities, and are not affected by desorption and drift, they may still be prevented from growing by the erosion or bouncing barriers which are discussed in context of this work in Section \ref{discussion}.

\begin{figure*}[!htb]
 \centering
 \includegraphics[width=15.8cm]{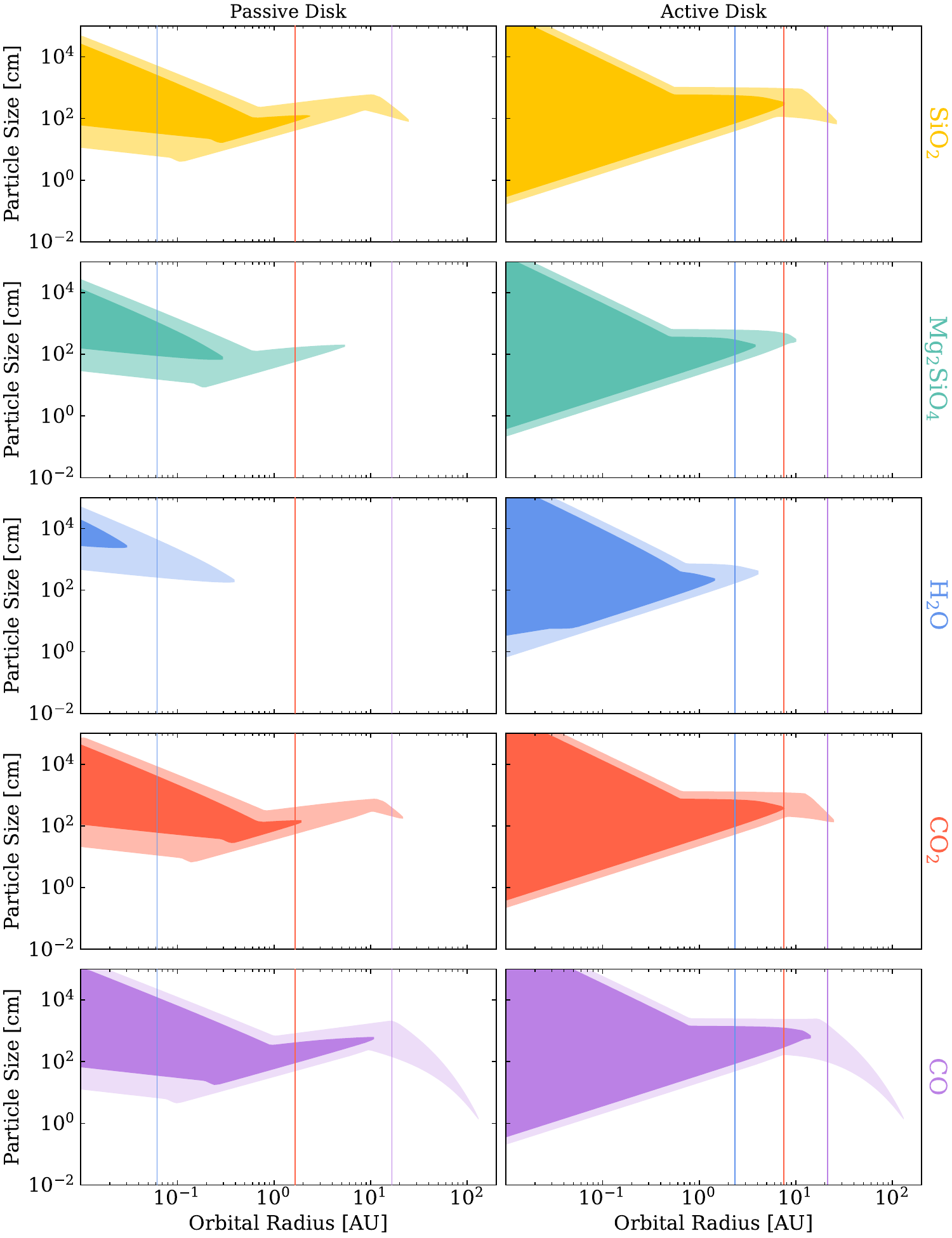}
 \caption{Fragmentation regions in particle size--orbital radius phase space for SiO$_2$ (yellow), Mg$_2$SiO$_4$ (green), H$_2$O (blue), CO$_2$ (red), and CO (purple) aggregate particles in TW Hya following the prescription from \cite{Wada2007,Wada2009b} (Eq. \ref{eq:vcrit_wada}). The left panels are for a passive disk and the right panels are for an active disk. The opaque region is for BPCA particles, while the fainter region is for BCCA particles. Ice line locations for H$_2$O (blue), CO$_2$ (red), and CO (purple) are shown as vertical lines. Regions for H$_2$O (third row) directly correspond to the black outline highlighted in the middle column of Figure \ref{fig:vrels_alpha}.}.
 \label{fig:frag_comp}
\end{figure*}

\begin{figure*}[!ht]
 \centering
 \includegraphics[width=17cm]{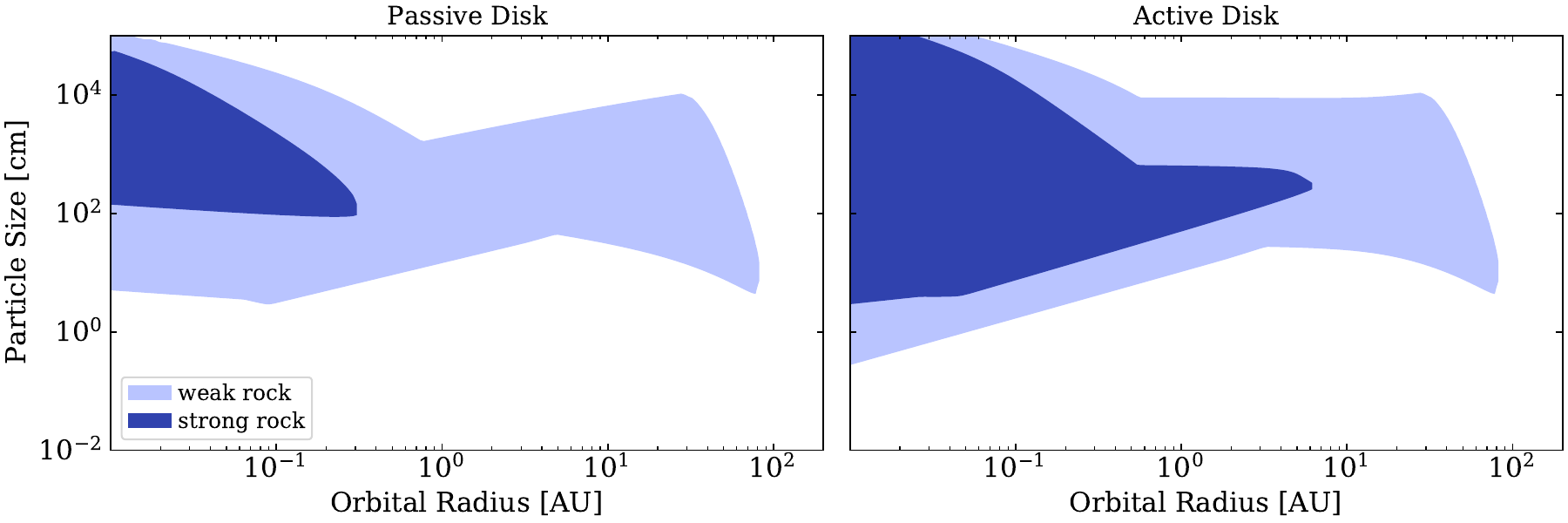}
 \caption{Fragmentation regions calculated for solid rocks using the prescription from  \cite{Stewart2009} (Eq. \ref{eq:vcrit_SL}), to be compared with the fragmentation regions from Figure \ref{fig:frag_comp}. Distinction between weak (light blue) and strong (dark blue) rock properties is discussed in Section \ref{sec:critical_velocity}. The left panels are for a passive disk and the right panels are for an active disk. These regions most closely resemble SiO$_2$ and Mg$_2$SiO$_4$ aggregate particles from Figure \ref{fig:frag_comp}. }.
 \label{fig:frag_SL}
\end{figure*}

While Figure \ref{fig:frag_comp} compares the fragmentation regions for a variety of compositions throughout the entire disk, particles composed of H$_2$O, CO$_2$, and CO ice are expected to desorb at their respective ice lines and are not expected to exist in the solid phase inward of their ice lines. The ice lines are represented as vertical lines with corresponding colors that match with composition color. The opaque colors represent BPCA particles and the more transparent colors represent BCCA particles. Silicate particles, as well as BCCA aggregates composed of CO$_2$ or CO ice, have the most widespread fragmentation regions although they are expected to be coated in ice beyond the H$_2$O ice line. Nearly all BPCA ices only have fragmentation regions within their respective ice lines, meaning that all compact particles---and grains which are coated in these ices---do not undergo collisional fragmentation. Desorption, and potentially other barriers as discussed in Section \ref{discussion}, will ultimately determine whether these particles will grow to large sizes. All growing silicate particles in the inner disk, within the H$_2$O ice line, are fragmenting. The silicates also match the expected fragmentation region for solid non-aggregate rocks using the critical disruption energy from \cite{Stewart2009} demonstrated in Figure \ref{fig:frag_SL}. 

%%%%%%%%%%%%%%%%%%%%%%%%%%%%%%%%%%%%%%%%%%%%%
%%%%%%%%%%%%%% GROWTH REGIONS %%%%%%%%%%%%%%%
%%%%%%%%%%%%%%%%%%%%%%%%%%%%%%%%%%%%%%%%%%%%%

\section{Favorable Regions for Planetesimal Growth}\label{favorable_regions}

We now focus on the interplay between collisional fragmentation and particle growth.  An initially-small particle grows via coagulation and drifts toward the star due to gas drag.  If growth and drift carry the particle into a fragmentation region, as illustrated in Figures \ref{fig:frag_comp} and \ref{fig:frag_SL}, fragmentation occurs, preventing growth to larger sizes.  If drift carries the particle past an ice line, desorption can cause it to lose the relevant ice species from its mantle and, potentially, to fall apart and begin growing anew in the absence of that solid species. We determine the fate of a particle by simultaneously integrating the equations for collisional growth, particle drift, and desorption, and then comparing our results with the fragmentation regions computed in Section~\ref{sec:frag regions} and ice line locations computed as described in Section~\ref{subsec: icelines}. 

\begin{figure}[!htb]
 \centering
 \includegraphics[width=8.4cm]{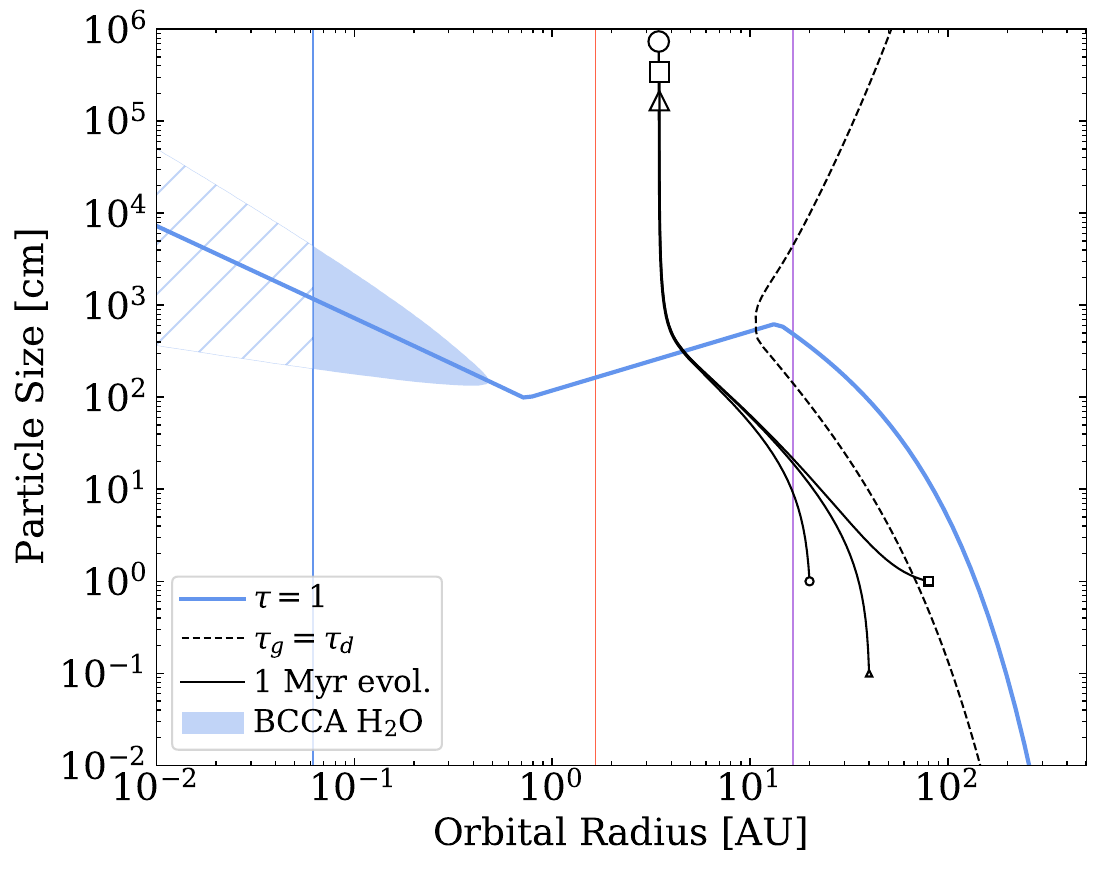}
 \caption{H$_2$O BCCA fragmentation region (shaded blue) for a passively heated disk with $f_{\rm d}=10^{-3}$, $\alpha=10^{-3}$, and $\phi_{\rm s}=0.3$. Fragmentation extends inwards of the H$_2$O ice line (barred blue), however the ice will sublimate and instead silicate fragmentation should be considered. The shape of the fragmentation region follows the H$_2$O $\tau=1$ line from Figure \ref{fig:TWHya_tau1} (blue line). The dashed black line is where the analytic growth and drift timescales (defined in Appendix \ref{A:timescales}) are equal.  After an initial phase of growth, particle evolution paths in the outer disk (solid black curves) follow the lower branch of this curve, offset by a multiplicative factor that arises from more accurate treatment of the integration. Each particle evolution (starting at the circle, square, and triangle) is shown for 1 Myr.  Note that all three converge to the same evolution path where the growth rate is balanced by the drift rate.}
 \label{fig:frag_explain}
\end{figure}

Figure \ref{fig:frag_explain} illustrates an example outcome for a BCCA particle composed of H$_2$O ice---including the $\tau=1$ profile (Figure~\ref{fig:TWHya_tau1}), the relevant fragmentation region (passive BCCA H$_2$O region from Figure~\ref{fig:frag_comp}), ice lines (derived in Section \ref{subsec: icelines}), and particle evolution paths (derived in Appendix~\ref{A:timescales}). The particle evolution paths are directly compared to regions of fragmentation, as well as ice lines, to check whether the particle will collisionally fragment or desorb during its evolution. 

Analytically, a particle is expected to grow and drift along a path where the coagulational-growth and particle-drift timescales are equal (Equations~\ref{eq:tgrow} and \ref{eq:tdrift}), which is similarly demonstrated in Appendix 1 of \cite{Tsukamoto2017} where they find that the growth and drift timescales eventually converge.  Numerically, we solve for the particle's mass and disk radial position over 1 Myr which is governed by the growth, drift, and desorption rates ($dm/dt$ and $dr/dt$ from Equation~\ref{eqn:coupledodes}). In Figure~\ref{fig:frag_explain}, the analytic solution is shown as a dotted black line, while the solid black lines are the numerical solutions to the coupled differential equations. The three numerical solutions are initialized with disk radii of $r=$ 20, 40, and 80 AU, with corresponding sizes of $s=$ 1, 0.1, and 1 cm (the different particles are represented with a circle, triangle, and square respectively). 

The general evolution of a particle in a disk includes the following phases: (1) the particle initially grows rapidly until the growth and drift timescales are comparable, (2) then the particle will grow and drift at the same rate until the particle's stopping time approaches unity ($\tau \sim 1$), and (3) once the particle reaches $\tau = 1$ at a certain orbital radius, the growth rate increases while the drift rate decreases causing the particle to grow in place. 

Phase (3), where growth becomes much faster than drift, is likely for porous aggregates, however particles of little porosity where $\phi_{\rm s} \rightarrow 1$ may evolve differently. \cite{Okuzumi2012} find that porous aggregates are not hindered by radial drift in the inner disk and can continue to grow whereas compact solid particles ($\phi_{\rm s} = 1$) are hindered, although they do not consider fragmentation. In all models, radial drift limits particle growth in the outer disk. We assume that particles are porous, keeping the filling factor the same to emphasize how particle structure and composition affect fragmentation. We use a value of $\phi_{\rm s} = 0.3$, for both BPCA and BCCA aggregates, to represent particles in protoplanetary disks \citep{Zhang2023}. BCCA particles will typically have lower filling factors than 0.3 depending on the particle size and number of fractal dimensions as demonstrated in \cite{Tazaki2019}. At lower particle filling factors, particle relative velocities are larger, causing fragmentation to be more likely (see Appendix \ref{A: size ratio and porosity}, Figure \ref{fig:vrels_sizeratio}). During phase (2), as particles are growing towards $\tau \sim 1$, collisions will have likely compactified (or fragmented) BCCA aggregates, making BPCA aggregates more relevant for phase (3).

While phase (2) can be well described by the analytic solution, the numerical solution is needed to accurately model particle behavior once $\tau\sim1$ in order to trace the desorption of an ice particle if it drifts interior to it's ice line. The particle will cease growing if it evolves into the corresponding species fragmentation region, but not necessarily if it drifts interior to its corresponding ice line. Since the particle may be composed of multiple species, the particle may fall apart at the ice line, however the smaller fragments will continue to grow and drift through the same phases described above.  Because growth curves converge to a path along which the growth and drift timescales are comparable, the extent to which particles are disrupted at ice lines before reforming with a new mantle composition (for example the CO ice line shown in purple in Figure~\ref{fig:frag_explain}) does not substantially affect our results.    In the H$_2$O BCCA example of Figure~\ref{fig:frag_explain}, neither fragmentation nor desorption prohibit particle growth indicating that the outer disk may be a region where collisional growth is favorable. 

As the particles drift inwards the overall dust-to-gas mass ratio, or the ratio between the solid and gaseous disk surface densities, will also evolve. At early disk lifetimes the dust-to-gas ratio is expected to be around $10^{-2}$ corresponding to ISM values \citep{Bohlin1978}. At later disk lifetimes once the disk has had time to evolve, the dust-to-gas ratio can decrease by an order-of-magnitude to a value of around $10^{-3}$, or even lower \citep[e.g.,][]{Birnstiel2010,Birnstiel2012,Powell2019a}. For our modeling purposes we keep the dust-to-gas ratio constant throughout the disk and with time even though it varies for both \citep{Alexander2007}. The varying $f_{\rm d}$ is relevant for the growth timescale (see Equation \ref{eq:tgrow}), thus we model particle evolution for both of these values to evaluate how composition-based fragmentation regions can change within a disk lifetime. 

\begin{figure*}[!htb]
 \centering
 \includegraphics[width=17cm]{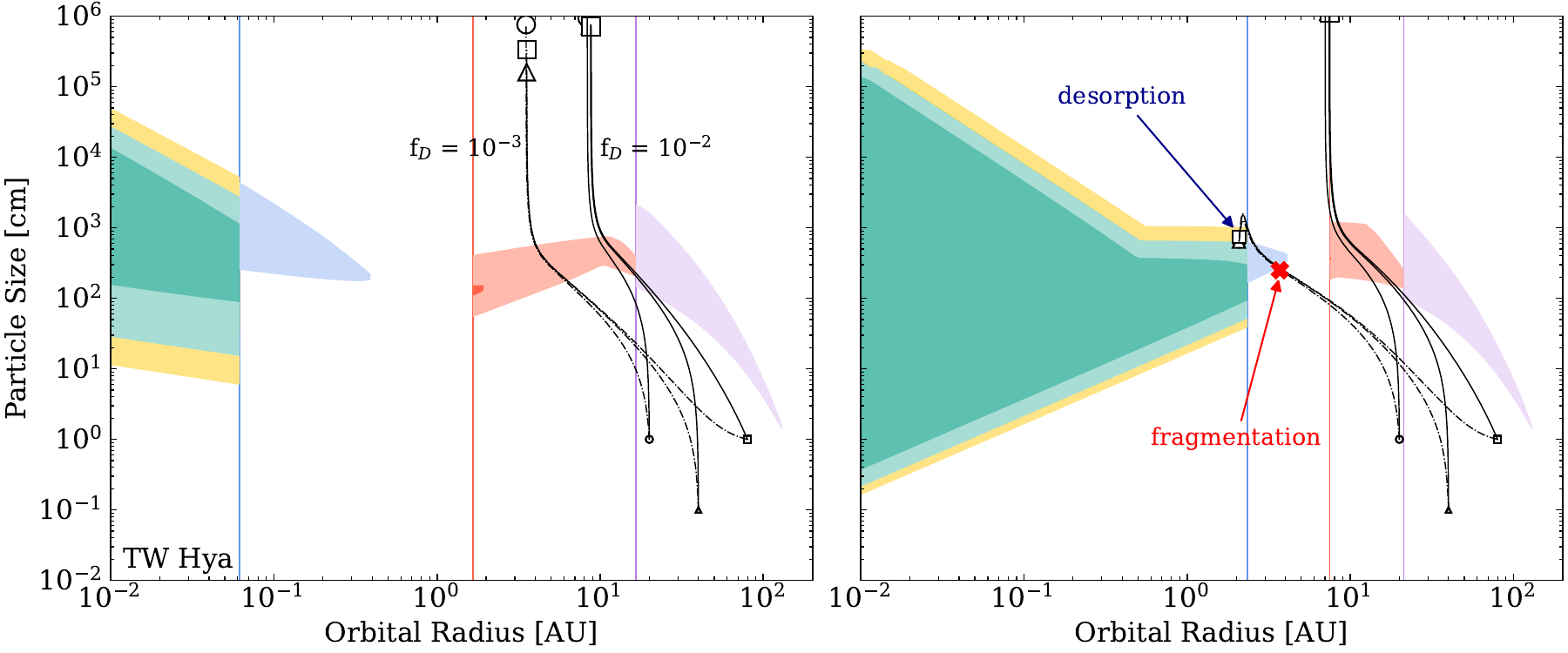} 
 
 \vspace{0.5cm}
 
 \includegraphics[width=17cm]{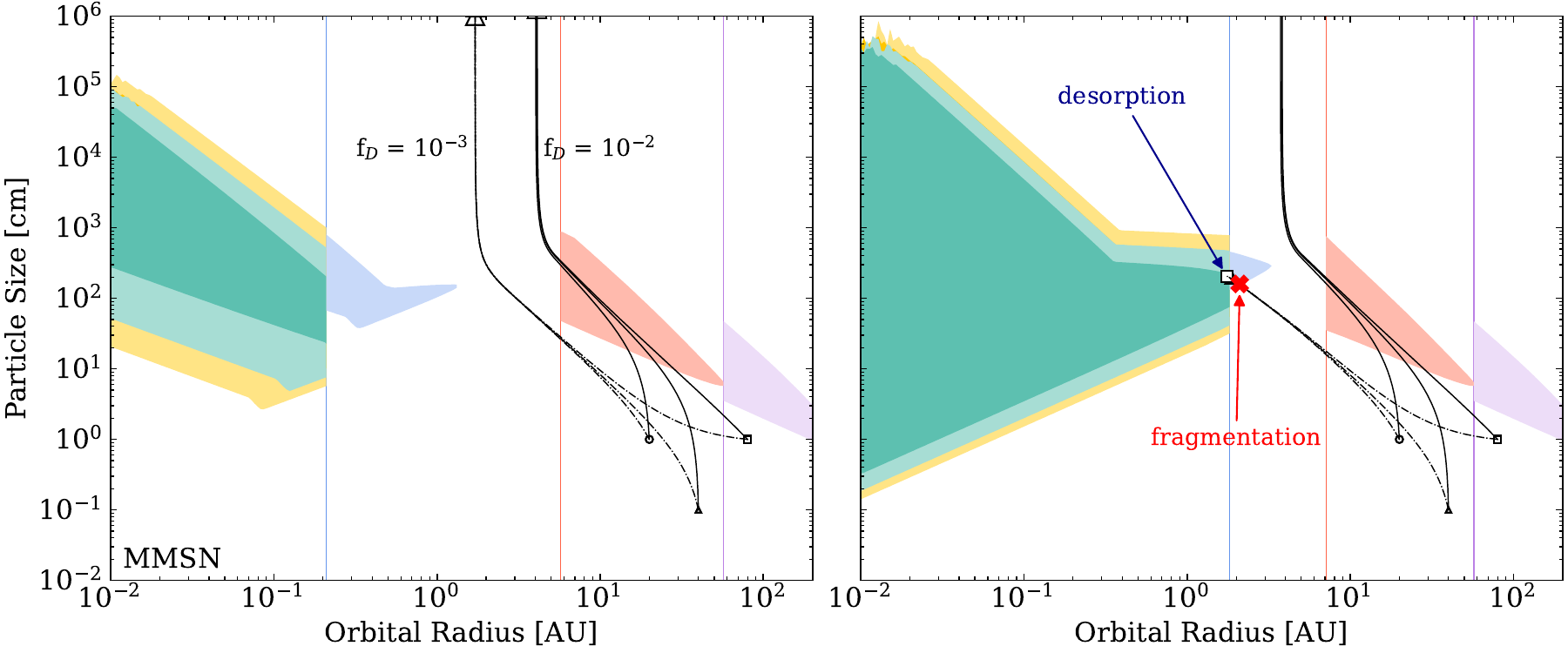} 
 \caption{Composite fragmentation regions with desorption of ices taken into consideration for TW Hya (top) and the MMSN (bottom)---leaving SiO$_2$, Mg$_2$SiO$_4$, BCCA H$_2$O, BCCA CO$_2$, and BCCA CO fragmentation. The left panels are for a passive disk and the right panels are for an active disk.} The black lines are numerically evolved particle evolution paths over 1 Myr for three H$_2$O ice particles which begin at a certain size and orbital radius: circle (1 cm, 20 AU), triangle (0.1 cm, 40 AU), square (1 cm, 80 AU). The solid lines are calculated with a dust-to-gas ratio of $f_{\rm d}=10^{-2}$ and the dotted-dashed lines are for $f_{\rm d}=10^{-3}$. The $f_{\rm d}=10^{-3}$ case for the active disk fragments and desorbs before growing to large sizes. Fragmentation is pointed out as the red cross. The other particle evolution paths, however, do not undergo fragmentation or desorption and are able to grow to sizes no longer affected by the meter-size barriers.
 \label{fig:frag_all}
\end{figure*}

With all of these pieces, the final component of our model incorporates all of the species (SiO$_2$, Mg$_2$SiO$_4$, H$_2$O, CO$_2$, and CO) into one fragmentation picture. In the inner disk, interior to the H$_2$O ice line, the fragmentation regions which dominate are those for SiO$_2$ and Mg$_2$SiO$_4$. Beyond the H$_2$O ice line, silicates are coated in ice (see Section \ref{sec:materialprops}), and the dominating fragmentation region is determined by the composition of the outermost layer of the ice mantle, monomer size, and how compact the aggregate is. Regions dominated by H$_2$O, CO$_2$, and CO are cut off inwards of their ice lines since these species are expected to desorb at the orbital radii of their ice lines leaving behind a silicate grain. Collisional fragmentation is not likely to occur for ice-coated BPCA particles as these fragmentation regions are mostly interior to their ice lines as illustrated in Figure~\ref{fig:frag_all}, which also includes the numerical evolution paths for growing particles. BCCA particles, however, are likely to undergo fragmentation. While BCCA aggregate collisions may result in compaction at small sizes, as they get larger their growth will be limited by the meter-size barriers.

The results of Figure~\ref{fig:frag_all} demonstrate that there do exist regions of the disk where growth can be efficient and will not be impeded by fragmentation or desorption depending on the assumed midplane temperature. In general, for a passively-heated disk, growth can efficiently happen at early and later disk lifetimes---all three H$_2$O particles evolve to sizes which are no longer limited by fragmentation. For an actively heated disk, which has a hotter midplane, at later times ($f_{\rm d}=10^{-3}$), the H$_2$O ice line is pushed further out in the disk such that particles will fragment before growing to large sizes.  In either case however, BPCA particles starting beyond the CO$_2$ ice line may experience efficient runaway collisional growth beyond the fragmentation and drift barriers for $f_{\rm d}=10^{-2}$. Based on particle evolution paths in context of fragmentation and desorption, the favorable growth region in TW Hya is between the CO$_2$ and CO ice lines, while for the MMSN it is between the H$_2$O and CO$_2$ ice lines. The differing regions between the MMSN and TW Hya are indicative that disk properties are an important factor in shaping favorable growth and fragmentation regions.

%%%%%%%%%%%%%%%%%%%%%%%%%%%%%%%%%%%%%%%%%%%%%
%%%%%%%%%%%%%% Discussion %%%%%%%%%%%%%%%
%%%%%%%%%%%%%%%%%%%%%%%%%%%%%%%%%%%%%%%%%%%%%

\section{Discussion}\label{discussion}

\begin{figure*}[!htb] 
 \centering
 \includegraphics[width=17.5cm]{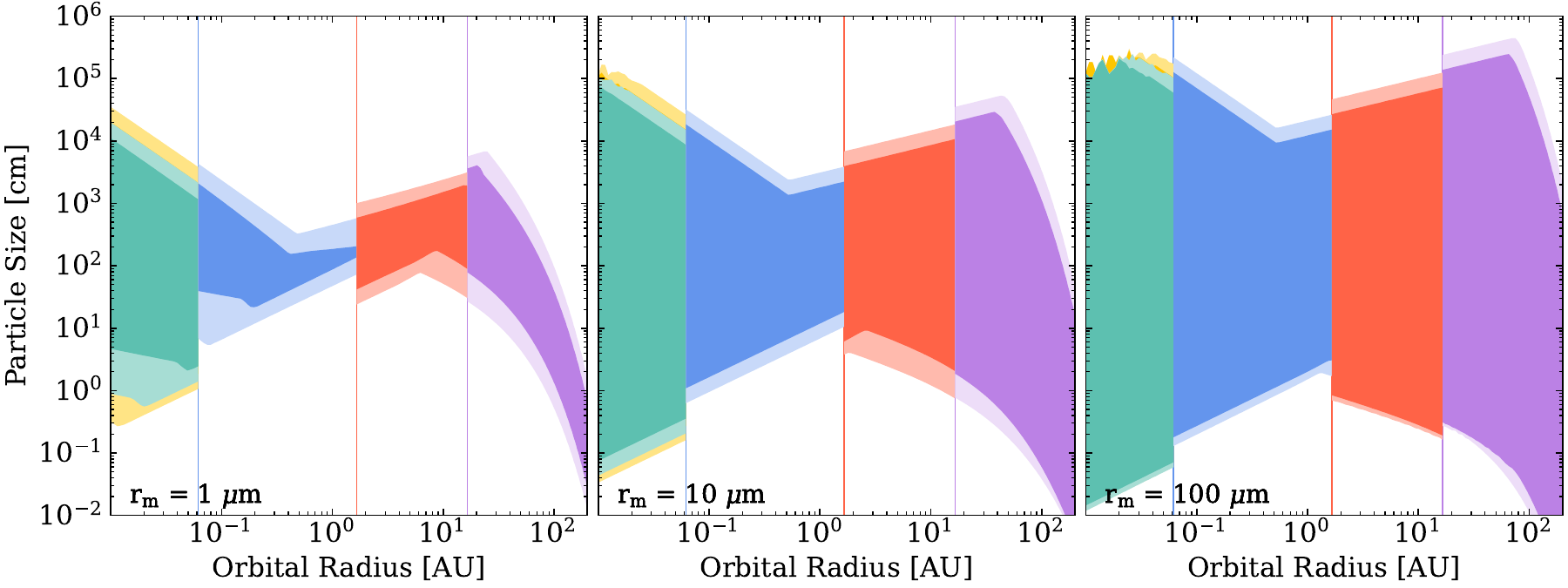}
 \caption{Aggregates composed of large monomers (1, 10 and 100 $\mu$m) have larger regions of fragmentation that can span the entire disk over many sizes for all particle compositions. As in Figures \ref{fig:frag_comp} and \ref{fig:frag_all}, the lighter and darker regions are BCCA and BPCA particles respectively, while colors correspond as yellow for SiO$_2$, green for Mg$_2$SiO$_4$, blue for H$_2$O, red for CO$_2$, and purple for CO. These Figures are for a passively heated TW Hya with $f_{\rm d}=10^{-3}$, $\phi_{\rm s}=0.3$, $\chi$=0.5, and $\alpha=10^{-4}$.}
 \label{fig:varymonomers}
\end{figure*}

In this work, we develop a flexible framework for particle evolution that can be updated readily with improved constraints on fragmentation from future observations, laboratory experiments, and theoretical studies of velocity-dependant growth barriers. In this Section we discuss model sensitivities, other potential barriers to particle growth, observational tests and implications, and necessary constraints from laboratory data that could be used to validate or improve the reliability of this modeling framework.

\subsection{Sensitivity to Monomer Size for Aggregate Fragmentation} \label{sec:mon_size_dependnece}

As illustrated in Figure~\ref{fig:monomer_fragmentation_dependance}, the critical velocity for destruction of aggregate particles is strongly dependent on the size of their component monomers, $r_{\rm m}$.  Our fiducial $r_{\rm m} = 0.1 \mu$m is comparable to the sizes of ISM grains and is hence a minimum reasonable value \citep{Oberg2021}, yielding a maximum critical fragmentation velocity.  In Figure~\ref{fig:varymonomers}, we recreate the top left panel of Figure~\ref{fig:frag_all} (TW Tya passive disk with BPCA and BCCA aggregate particles and $\alpha = 10^{-4}$) for differing monomer sizes $r_{\rm m} = 1$, 10, and 100$\mu$m. Micron-sized monomers may narrowly find favorable growth regions within a disk in the case of BPCA aggregates, but for larger monomer sizes of 10 and 100$\mu$m no favorable regions exist.

As discussed in Section~\ref{sec:materialprops}, monomer sizes are uncertain, and indeed may vary throughout the protoplanetary disk lifetime.  In particular, very early in the disk lifetime, significant supersaturation of condensible gases likely allows the largest particles to accumulate substantial ice mantles via adsorption \citep{Powell2022}.  Our choice of small monomers is most likely to apply after condensible volatiles are depleted by the growth and drift of these ice-rich particles.  Because the surfaces of larger particles have less curvature, they nucleate condensation of volatiles more effectively (a process known as the Kelvin effect), meaning that small grains can remain ice-free even at early times \citep{Powell2022}.  Hence, a time-evolving exploration of appropriate monomer sizes requires a complete size distribution of particles rather than the consideration of typical particle sizes applied in this work.  Given our results, future work is merited to explore appropriate monomer sizes.  This work will require self-consistent computation of the evolution of disk volatile abundances alongside particle coagulation, compactification, and volatile adsorption over the full distribution of particle sizes.  

In the absence of such a full model, we appeal to Figure 1(b) of \citet{Powell2022} to demonstrate why we consider our fiducial choice to be plausible. The figure shows the abundances of ice-free and ice-coated particles at 30AU in TW Hya as a function of particle size at a late disk age of 5 Myr. Particles grow ice-free from $s \sim$0.1-100$\mu$m (first stage of Figure~\ref{fig:growth_cartoon}).  At these small sizes, relative velocities are small (see Figure~\ref{fig:vrels_alpha}) and collisional growth likely leads to production of fluffy aggregates \citep[e.g.,][]{Smirnov1990,Meakin1991,Kempf1999,Blum2000a,Krause2004,Wada2009b,Paszun2006,Kataoka2013}.
As particles approach $s\sim 1$mm, they become coated in ice and then drift inward.  Particles that accreted ice early have already drifted away. The illustrated icy particles have adsorbed moderate ice coatings on top of aggregates having monomers of the initial small particle size, taken to be 0.1$\mu$m inherited from the ISM.  These icy particles are consistent with case (a) of Figure~\ref{fig:growth_cartoon}.  

\subsection{Other Potential Barriers to Growth}\label{sec:disc_other_barriers}
Fragmentation is one of the most significant barriers to early stages of planet formation, however, there may be other barriers to continued small-particle growth, such as the bouncing and aeolian-erosion barriers. We note that the velocity of collisions where the bouncing barrier is relevant is not yet well constrained and may strongly depend on particle properties \citep[e.g.,][]{Blum1993,Langkowski2008,Windmark2012}.

\cite{Nietiadi2020} find that for small particles covered in ice mantles, the bouncing barrier critical velocities range from 10 - 100 m s$^{-1}$ and are likely to increase for particles of larger size, meaning that the bouncing barrier limits collisional growth at higher relative velocities than the fragmentation velocities. \cite{Sirono2017} find that sintered BPCA H$_2$O ice particles are likely growth-limited by the bouncing barrier rather than the fragmentation barrier as the monomer connections are stronger. They also find that sintered BCCA aggregates become more susceptible to fragmentation as compared to the unsintered case. 
Sintered particles are likely to exist within the disk environment, but for our purposes we only consider the unsintered case since the critical velocities from \cite{Sirono2017} fall within the critical relative velocity range in our study. Future work should include a distinction between the differing critical displacements for breaking monomer connections within a sintered versus unsintered aggregate particle. The work presented here may be unaffected by sintering and the bouncing barrier as the ice particles that survive and undergo runaway growth never reach sufficiently high relative velocities where the bouncing or fragmentation barrier are efficient in limiting particle growth. Improved constraints on the bouncing barrier are necessary for a thorough understanding of particle evolution in disks. 

The aeolian-erosion barrier \citep[e.g.,][]{Paraskov2006,Schrapler2011,Rozner2020} can efficiently erode larger pebbles and boulders, ranging from $\sim 10-1000$ meters, down to centimeter sizes. Similar to collisional fragmentation, the aeolian-erosion barrier depends on a grain size (comparable to the monomer size in our discussion of collisional fragmentation), and the species's surface energy. \citet{Rozner2020} show that when grain sizes are 0.1 cm, the aeolian-erosion barrier can halt particle growth in disks at $\sim$10 cm. However, if we assume that the grain size is comparable to the nominal monomer size considered in this work (0.1 $\mu$m), the velocity threshold for the onset of aeolian-erosion increases by a factor of 100 such that aeolian-erosion will not significantly limit growth in protoplanetary disks. Future work may expand the model to include the aeolian-erosion barrier in order to evaluate the interplay between growth, drift, and aeolian-erosion for a variety of monomer sizes during early stages of planet formation. 

In general, a complete dynamical model of particle evolution should include growth, drift, sublimation, and fragmentation, as well as other potential barriers to growth. While this is beyond the scope of our work, we highlight their importance for future studies.

\subsection{Disk Substructure and Streaming Instability}

Sustained particle growth in the favorable growth regions may potentially give rise to millimeter emission substructure. Rapidly growing particles in the disk-dependant favorable growth region (between H$_2$O and CO$_2$ ice lines for the MMSN, and between CO$_2$ and CO ice lines for TW Hya) can reach large sizes without drifting inwards, thus causing a pileup of particles resulting in a dearth of millimeter emission just interior to the favorable growth region. This may result in gaps in emission that would be present even in the case of young disks with larger dust-to-gas ratios \citep[e.g.,][]{Williams2011,Andrews2015}. Several disks presented in \cite{Huang2018} from the DSHARP ALMA survey have significant substructure near their respective CO$_2$ ice line. Observations often look at older unobscured disks ($f_{\rm d}=10^{-3}$), however, our results demonstrate that even at younger times particles can overcome the fragmentation barrier. Thus the work presented here may also be able to explain how significant millimeter substructure arises in young disks such as HL Tau \citep{ALMAPartnership2015}. 
Furthermore, If favorable growth regions are preventing particles from drifting interior to the H$_2$O ice line, the abundance of gaseous H$_2$O in the inner disk will be depleted. This H$_2$O depletion is seen in several disk observations with the Herschel Space Observatory \citep[e.g.,][]{Bergin2010,Bergin2013,Kamp2013,Du2015,Salinas2016,Du2017}, and may be explained with the favorable growth regions presented in our work.

Favorable growth regions are locations in the disk where particles can pileup as a result of their drift timescale becoming much longer than their growth timescale. We note that these regions will have higher dust-to-gas ratios, making them favorable for not only collisional growth but also for planetesimal formation through direct gravitational collapse. In particular, these pileups of growing particles could potentially instigate Resonant Drag Instabilities, particularly because particles in the favorable growth region naturally have $\tau \sim 1$, the approximate particle size most favorable to these instabilities and have higher densities of dusty and icy material.

\subsection{Laboratory Constraints}

Physical processes governing the growth and fragmentation of particles in disks strongly depend on laboratory derived material properties. In particular, a species surface energy, which is typically not well-constrained, determines the fragmentation properties of particles composed of that species. Improved surface energy constraints for all potential refractory and volatile species present in disks are thus necessary for an accurate picture of particle fragmentation. Furthermore, the surface processes during particle impacts are also not yet well-constrained. In this work, we have assumed dry surface physics, however, collision-induced heating can potentially melt the ice on a particle's surface which can decrease the likelihood of particle fragmentation \citep{Nietiadi2020}. This may resemble sintered aggregates, although more work needs to be done in distinguishing the various surface connections, especially for particles coated in a variety or mixture of ices \citep{Sirono2017}. Disk particles with mixed compositions may also have collisional properties that vary from those presented in this work due to differing structures between the grain--ice and ice--ice layers \citep[e.g.,][]{Fogarty2010}. For example, when \cite{Musiolik2016b} tested mixtures of ice (CO$_2$ + H$_2$O) the experiments found an order-of-magnitude increase in sticking velocity, indicating that ice mixtures can result in decreased particle fragmentation. Future work that describes the surface and collisional properties of mixed and partially-melted ice particles with a variety of disk-relevant compositions will be useful in improving models of particle evolution in disks.

\section{Summary \& Conclusions} \label{summary}

We develop a particle growth, drift, and fragmentation model which self-consistently solves for a particles' stopping time and includes ice sublimation, to test how different particle compositions and properties can withstand fragmentation. We find that ice particles in the outer disk, in between the CO$_2$ and CO ice lines for TW Hya and in between the H$_2$O and CO$_2$ ice lines for the MMSN, may efficiently grow to large sizes without fragmenting or sublimating.

Our model produces regions of fragmentation in particle size--disk orbital radius phase space by comparing a particles' relative velocity with material and structural dependant critical fragmentation velocities. We test our results by varying the turbulence ($\alpha$), dust-to-gas ratio ($f_{\rm d}$), size ratio between colliding particles ($\chi$), particle filling factor ($\phi_{\rm s}$), monomer size ($r_{\rm m}$), disk temperature profile, and in particular particle composition. This study evaluates the fragmentation regions of the refractory species SiO$_2$ and Mg$_2$SiO$_4$, the volatile species H$_2$O, CO$_2$, and CO in their ice form, as well as a more general prescription for weak and strong non-aggregate rock.

For the fiducial case of TW Hya ($\alpha=10^{-3}$, $\chi=0.5$, $\phi_{\rm s}=0.3$, $r_{\rm m}$ = 0.1 $\mu$m) our model shows the following behavior:

\begin{itemize}
    \item In passive disks, all compact BPCA ice-coated H$_2$O, CO$_2$, and CO aggregate particles beyond $\sim 1$AU are able to grow without fragmenting or desorbing. This is the case for both the MMSN and TW Hya, and is independent of disk age.

    \item In active disks, and at early times when $f_{\rm d}=10^{-2}$, compact BPCA ice-coated aggregate particles are also able to grow unimpeded beyond $\sim 10$AU.

    \item Fluffier BCCA aggregate particles have fragmentation regions present throughout the disk making such particles more susceptible to fragmentation as compared to growth. BCCA H$_2$O ice particles, however, are still able to grow beyond the meter-size in all scenarios except at late times ($f_{\rm d}=10^{-3}$) in active disks.

    \item Uncoated silicate particles are still expected to collisionally fragment throughout the disk. In particular, SiO$_2$ using the adapted critical velocity from \cite{Wada2007,Wada2009b} and weak rock using the adapted formalism from \cite{Stewart2009} will always have relative velocities reaching critical fragmentation velocities. Mg$_2$SiO$_4$ and strong rock are less susceptible to fragmentation, and have fragmentation regions which reach to $\sim$1 AU and $\sim$10 AU for passive and active disks respectively.
 
    \item The critical fragmentation velocity decreases and particles become more susceptible to fragmentation as turbulence increases. H$_2$O ice will fragment throughout the disk with $\alpha=10^{-2}$. While for a passive TW Hya with $\alpha=10^{-4}$, H$_2$O ice will never reach collisional fragmentation velocities. 
 
    \item{Aggregates composed of larger monomers ($r_{\rm m}>10$$\mu$m as opposed to 0.1$\mu$m) will likely undergo fragmentation.}
 
\end{itemize}

These results indicate that particle growth may happen efficiently through collisions, without fragmenting or sublimating, beyond the H$_2$O ice line for the MMSN and beyond the CO$_2$ ice line for TW Hya.

\begin{acknowledgments}
D.P. acknowledges support from NASA (the National Aeronautics and Space Administration) through the NASA Hubble Fellowship grant HST-HF2-51490.001-A awarded by the Space Telescope Science Institute, which is operated by the Association of Universities for Research in Astronomy, Inc., for NASA, under contract NAS5-26555. R.M.C. acknowledges support from NSF CAREER grant number AST-1555385 and support from NASA Interdisciplinary Consortia for Astrobiology Research (ICAR) grant 80NSSC21K0597. 

\smallskip
 
E.S.Y. would like to thank Karin Öberg for insightful discussions, and is grateful to the mentors, family, and friends who have supported and encouraged me throughout this work. We also thank the anonymous referee for their very helpful and thorough review.  

\end{acknowledgments}

\bibliography{bibliography}{}
\bibliographystyle{aasjournal}

\appendix
\section{Deriving the Reynolds number for Distinguishing Between the Stokes and Ram Pressure Drag Regimes} \label{re derivation}

\begin{figure*}[!htb]
 \centering
 \includegraphics[width=13cm]{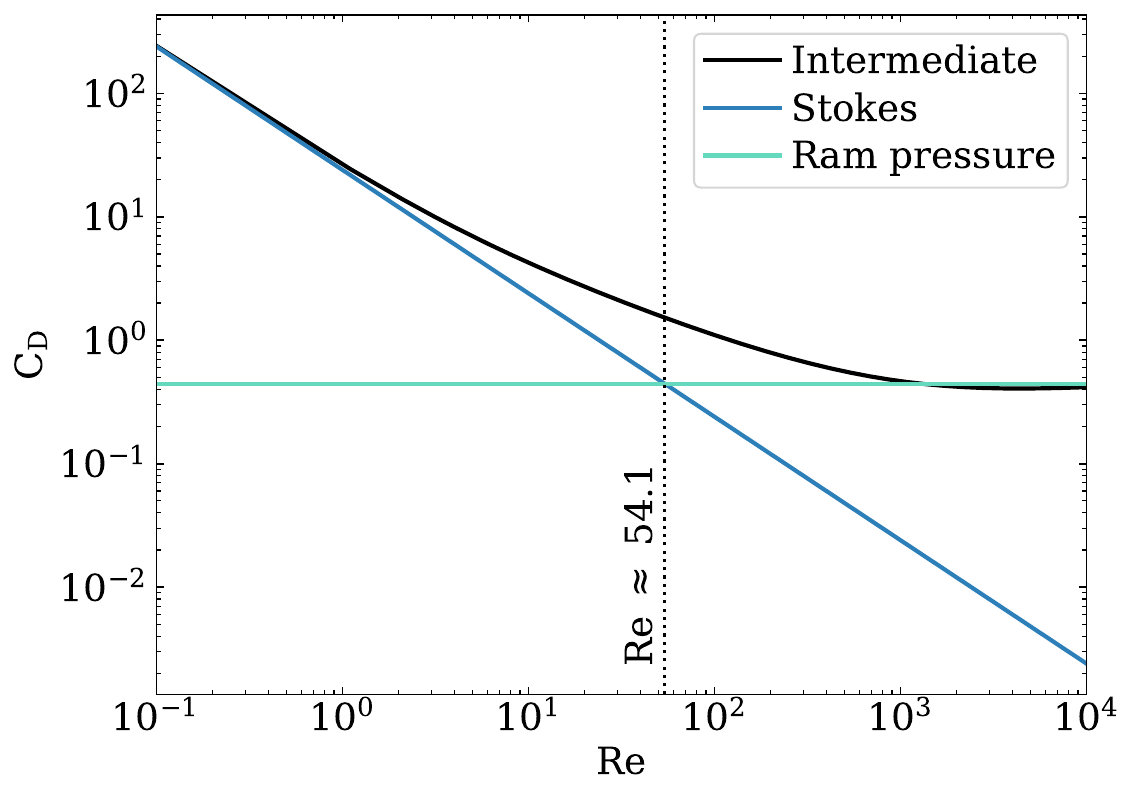}
 \caption{Comparison of the analytic expressions of the coefficients of drag $C_{\mathrm{D}}$ for the intermediate, Stokes, and ram pressure drag laws. The vertical dotted line is the switch between Stokes and ram pressure regimes, and is found to be at Re $\sim 54.1.$}.
 \label{fig:Re}
\end{figure*}

For our modeling purposes, we do not include the intermediate gas drag regime which spans Reynolds numbers $1<\mathrm{Re}<800$, in between the Stokes and ram pressure regimes. Instead, we extend the Stokes and ram pressure regimes into this Re phase space and use their intersection as the transition between the two. In general, the drag force is $F_{\mathrm{D}} = 0.5C_{\mathrm{D}}\pi r^2\rho_{\mathrm{g}} v_{\mathrm{rel}}^2$ \citep{brown2003,cheng2009,Perets2011}, where $C_{\mathrm{D}}$ is the coefficient of drag. For $s > (9/4)\lambda$, the Stokes and ram regimes have
\begin{equation}
 \begin{gathered}
  C_{\mathrm{D,Stokes}}={\frac{24}{\mathrm{Re}}},\\
  C_{\mathrm{D,ram}} = 0.44 \;\;.
 \end{gathered}
\end{equation}
The intermediate regime between them can be modeled using the \cite{cheng2009} fitting function 
\begin{equation}
  C_{\mathrm{D,intermediate}}={\frac{24}{\mathrm{Re}}\left(1+0.27\mathrm{Re}\right)^{0.43} + 0.47\left[1-\exp\left(-0.44\mathrm{Re}^{0.38}\right)\right]}.
\end{equation}
Figure~\ref{fig:Re} compares the $C_{\mathrm{D}}$ in each regime as a function of Re. The intersection between Stokes and ram pressure occurs at $\mathrm{Re}\sim54.1$, where $C_{\mathrm{D}}$ is only smaller by a factor of $\sim4.25$ from the intermediate value.  

\newpage

\section{Relative Velocity dependencies on Size Ratio and Filling Factor} \label{A: size ratio and porosity}
\begin{figure*}[!htb]
 \centering
 \includegraphics[width=17cm]{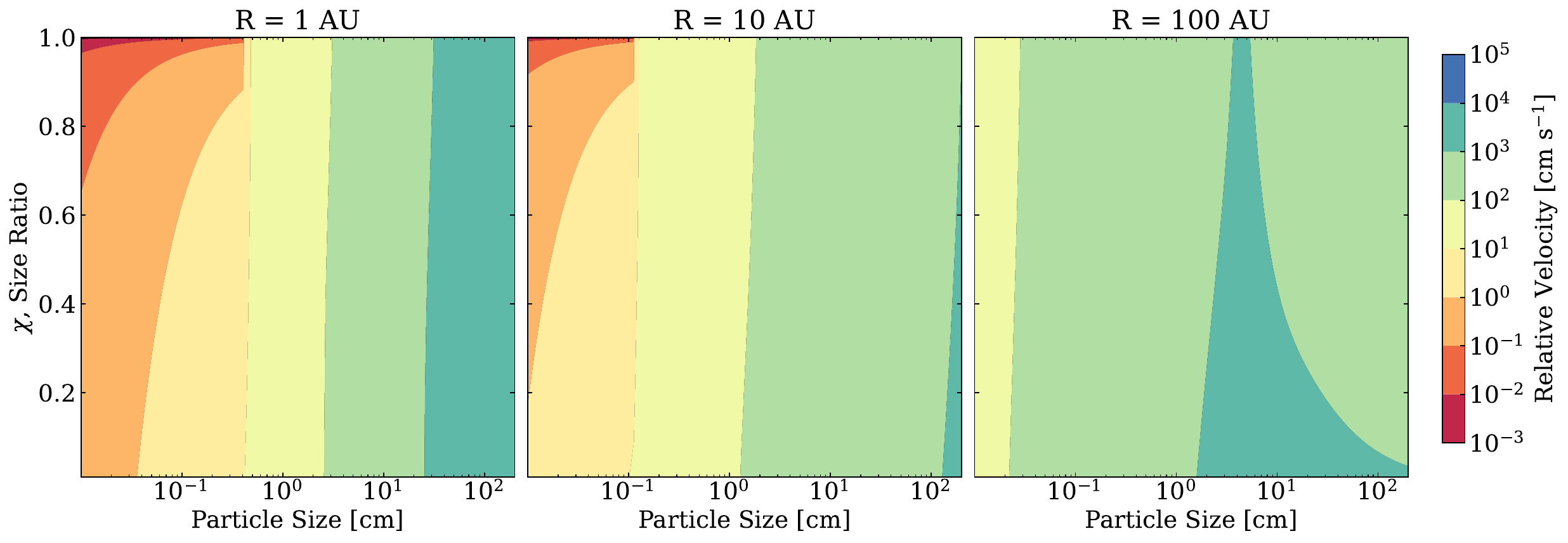}
 \includegraphics[width=17cm]{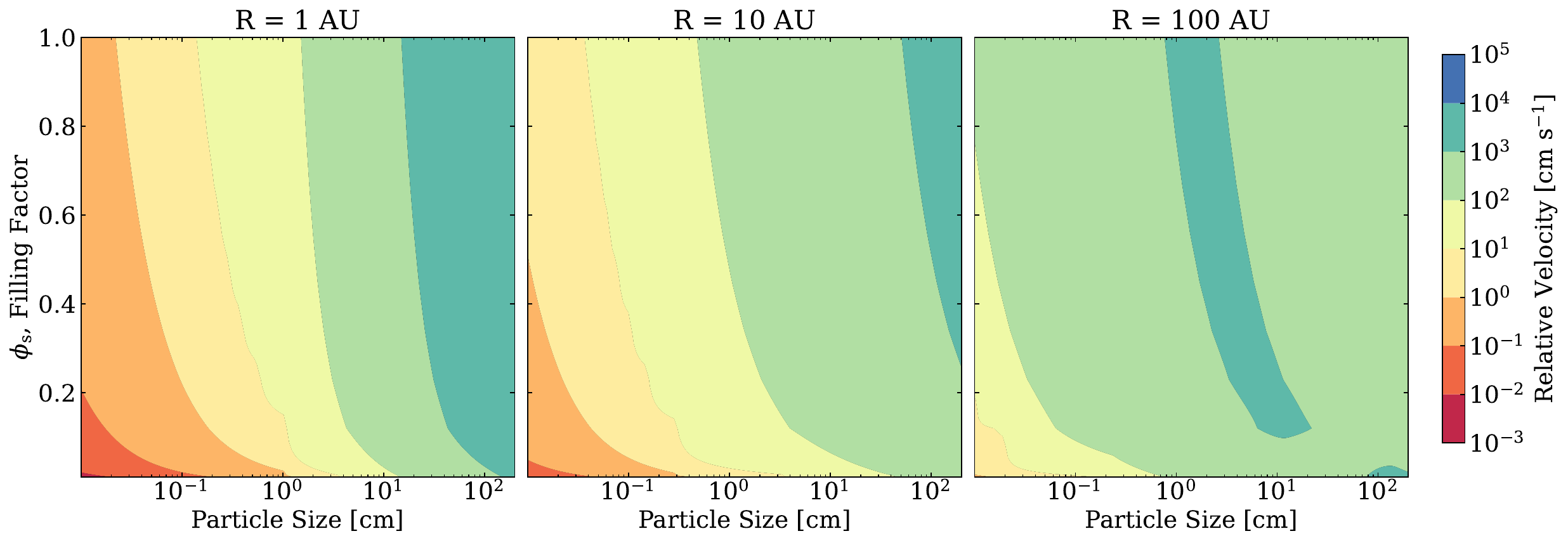} 
 \caption{Size ratio ($\chi$) and filling factor ($\phi_{\rm s}$) do not affect the relative velocity significantly for particles larger than mm-sizes beyond the H$_2$O ice line. These velocities are calculated for particles in a passive TW Hya with $f_{\rm d}=10^{-3}$ and $\alpha=10^{-3}$. Each panel from left to right is a disk radial snapshot at 1, 10, and 100 AU. These dependencies are nearly the same for an actively heated disk.}.
 \label{fig:vrels_sizeratio}
\end{figure*}

This appendix focuses on the effects of size ratio ($\chi$) and filling factor ($\phi_{\rm s}$, where porosity is $1-\phi_{\rm s}$) on the particle--particle relative velocity ($v_{\mathrm{rel}}$) calculation at 1, 10 and 100 AU. Collisions between two particles are likely to occur for a target particle with a projectile particle that are not the same size. Specifically, our model defines the projectile radius as $s_{\mathrm{p}} = \chi s_{\mathrm{t}}$ where $s_{\mathrm{t}}$ is the target radius, and the size ratio throughout the entire model is then $\chi \equiv s_{\mathrm{p}}/s_{\mathrm{t}}$.  We define the particle density $\rho_{\mathrm{s}} = \phi_{\rm s}\rho_{\mathrm{i}}$, where $\rho_{\mathrm{i}}$ is the material density (Table\ref{material_props}) and $\phi_{\rm s}$ is the filling factor. Figure~\ref{fig:vrels_sizeratio} demonstrates the effects of varying $\chi$ and $\phi_{\rm s}$ on $v_{\mathrm{rel}}$.

\smallskip

For nearly all particle sizes of interest, $\chi$ does not have a strong affect on $v_{\rm rel}$. In the inner disk, the smallest particles are in the regime labeled ``tightly coupled" in Appendix A of \citet{Powell2019a} and $v_{\rm rel}$ is dominated by the gas velocity in such a way that as $\chi$ increases, $v_{\rm rel}$ decreases. However, the particle size increases and transitions into the ``intermediately coupled" regime at sizes well below $\tau = 1$, at which point $v_{\rm rel}$ becomes nearly independent of $\chi$. For typical particle evolution's, the onset of fragmentation---if it occurs---happens in the intermediately coupled regime, meaning that our results are insensitive to $\chi$.  The relative velocity remains fairly consistent with varying $\phi_{\rm s}$ for nearly all particle sizes and does not have any transitions due to gas coupling effects. Particles with $\phi_{\rm s} < 0.3$ begin to have higher $v_{\rm rel}$ and are more likely to fragment. We choose fiducial values of $\chi=0.5$ and $\phi_{\rm s}=0.3$ \citep{Tazaki2019,Zhang2023}, and note that these values can be readily changed upon model initialization.

\newpage

\section{Fragmentation Regions for the MMSN}\label{A: MMSN}
The following includes the compositional fragmentation regions calculated for the MMSN. Disk mass plays an important role in setting not only the ice line positions, but also the overall fragmentation profile throughout the disk. In comparison to the fragmentation regions calculated for TW Hya (see Figure \ref{fig:frag_comp}), the MMSN regions are somewhat slimmer in terms of fragmenting sizes and have the Epstein-Stokes transition at a smaller orbital radius. 

\begin{figure*}[!htb]
 \centering
 \includegraphics[width=14cm]{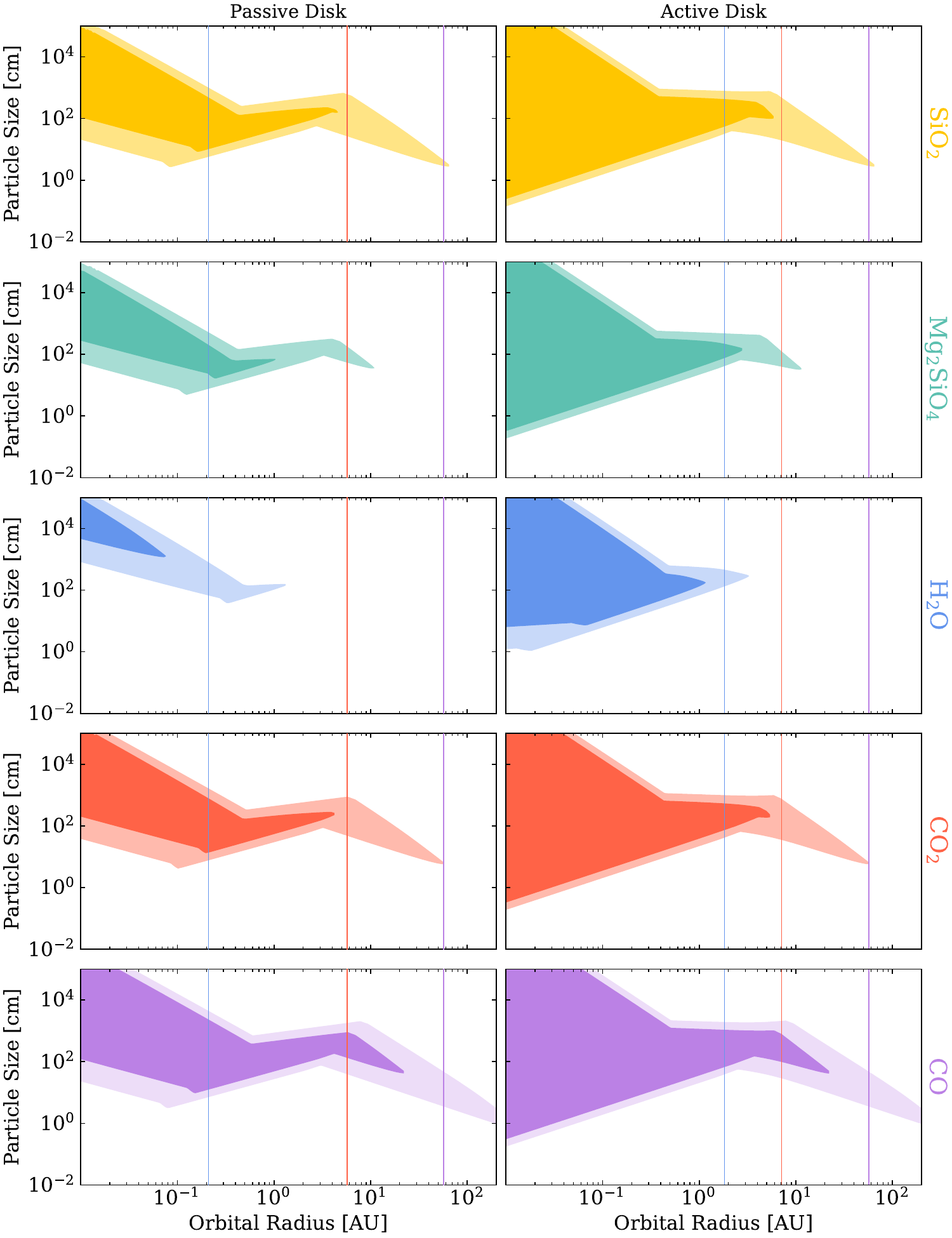}
 \caption{ Fragmentation regions in particle size--orbital radius phase space for aggregate particles of varying compositions in the MMSN. The left panels are for a passive disk and the right panels are for an active disk. The opaque region is for BPCA particles, while the fainter region is for BCCA particles. Ice line locations are shown as vertical lines with corresponding colors to compositions. }
 \label{fig:MMSN compositional frag}
\end{figure*}

\newpage

\section{Growth and Drift Rates and Timescales}\label{A:timescales}

The growth and drift timescales are used in calculating both the analytic and numerical particle evolution paths tracing their growth and inward drift (desorption is also included in particle evolution and is described in Section \ref{subsec: icelines}). The timescale definitions and derivations follow the works of \cite{Whipple1972}, \cite{ Weidenschilling1977a}, \cite{Birnstiel2012}, and \cite{Powell2019a}.

The collisional growth timescale is defined as $\tau_{g} = m/\dot{m}$, where $m=(4\pi/3) \rho_{\text{s}} s^3$ is the mass of the particle, and $\dot{m}=\rho_{\text{d}} \sigma v_{\mathrm{rel}}$ is the collisional growth rate. The radius of the largest particle is $s$, $\rho_{\mathrm{s}}=\phi_{\rm s}\rho_{\rm i}$ is the internal density of the particle which is dependent on filling factor and composition, and the cross sectional area of the particle is $\sigma = \pi s^2$. The volumetric mass density of the solid particles in the disk is simply $\rho_d = f_{\mathrm{d}}\rho_{\mathrm{g}}$, where $f_{\mathrm{d}}$ is the dust-to-gas ratio and $\rho_{\mathrm{g}}$ is the density of the gas. Combining everything together and including the terms mentioned above, we are left with the first expression in Eq.~(\ref{eq:tgrow}) which is then modified to the growth expression from \cite{Powell2019a}

\begin{equation}\label{eq:tgrow}
  \tau_{\text{g}} = \frac{4\rho_{\text{int}}s}{3f_{\text{d}}\rho_{\text{g}} v_{\mathrm{rel}}} \sim \frac{8s\rho_{\mathrm{int}}H_{\rm p}}{3f_{\rm d}\Sigma v_{\rm rel}f}
\end{equation}
where $f=0.55$ is a calibrating coagulation efficiency parameter, following \cite{Birnstiel2012}. We assume that the the particles perfectly stick together when they collide with other like-sized particles as long as the relative velocity does not surpass the critical fragmentation velocity \citep[see][]{Blum2008}.

The drift timescale is how long it takes a particle to drift inwards as it loses angular momentum due to gas drag (detailed in Section \ref{subsec:vrel_deriv}). The timescale is then $\tau_d = |r/\dot{r}|$, where $r$ is the orbital radius and $\dot{r}$ is the the radial drift velocity. The radial drift velocity is the same as the radial component of the relative laminar velocity from Equation~(\ref{eq.vr}). The drift timescale is then reduced to 
\begin{equation}\label{eq:tdrift}
  \tau_{\text{d}} = \frac{1}{2\eta \Omega}\left( \frac{1 + \tau^2}{\tau} \right),
\end{equation}
noting that this expression does not make the assumption that $\tau$ must always be less than 1. 

The growth and drift timescales play an important role in understanding which regions of the disk contribute to rapid growth or to rapid influx of materials. The sizes and orbital radii at which these timescales are equal is the general path expected for particle evolution in the outer disk. While this is an approximation, and differs slightly from the numerical calculation (see Figure~\ref{fig:frag_explain} for visual comparison between the two), it provides a good intuition for the regions of the disk in which particles are growing faster than drifting, or vice versa. We  numerically solve the set of differential equations below for the mass and orbital radius of the particle:
\begin{equation}\label{eqn:coupledodes} 
\begin{aligned}
\frac{dm(s)}{dt} &= \frac{m(s)}{\tau_{\rm g}} - 4\pi s^2 \mu_i F_{\rm desorp} \\
\frac{dr}{dt} &= -\frac{r}{\tau_{\rm d}}
\end{aligned}
\end{equation}

The mass is then converted into particle radius in order to directly compare the particle evolution path to the fragmentation regions. We use SciPy's general ordinary differential equation solver \texttt{integrate.odeint} \citep{Virtanen2020}.  Adsorption is not explicitly included in this calculation---the growth rate due to adsorption is significantly smaller than the collisional growth rate. The desorption rate is only important interior to the relevant ice line.

Initially particles will grow quickly until the growth and drift timescales become comparable (see Figure \ref{fig:frag_explain}). The particle will continue to evolve along the $\tau_{\rm g}=\tau_{\rm d}$ path until the particle becomes large enough such that $\tau\sim1$. Beyond this point, further growth causes the drift timescale to decrease and the particle will continue to grow at that location in the disk---typically at orbital radii between $\sim1-10$ AU. We note that because the drift timescale is shortest for $\tau = 1$, at larger disk radii than this location, $\tau_{\rm g} = \tau_{\rm d}$ yields two solutions, the smaller of which is relevant for growing particles. Interior to that disk radius, the growth timescale is always shorter than the drift timescale, and there are no solutions.

\end{document}